\documentclass{aa}  
\usepackage{graphicx}
\usepackage{txfonts}
\begin{document}
   \title{Colour-colour diagrams and extragalactic globular cluster ages}

   \subtitle{Systematic uncertainties using the $(V-K)-(V-I)$ diagram}

   \author{Maurizio Salaris
          \inst{1,2}
          \and
          Santi Cassisi\inst{3}}

   \offprints{Maurizio Salaris}

   \institute{Astrophysics Research Institute, Liverpool John Moores University,
              Twelve Quays House, Birkenhead, CH41 1LD, UK\\
              \email{ms@astro.livjm.ac.uk}
         \and
              Max-Planck-Institut f\"ur Astrophysik,
              Karl-Schwarzschild-Strasse 1, Garching, D-85748\\    
         \and
             INAF - Osservatorio Astronomico Collurania, Via Mentore
              Maggini, Teramo, I-64100\\
             \email{cassisi@oa-teramo.inaf.it}
             }

   \date{Received ; accepted}

 
  \abstract
   {Age and metallicity estimates for extragalactic globular
    clusters, from integrated colour-colour diagrams, are examined.}
   {We investigate biases in cluster ages and [Fe/H]
   estimated from the $(V-K)-(V-I)$ diagram, arising from   
   inconsistent Horizontal Branch morphology, metal mixture, treatment of core  
   convection between observed clusters and the theoretical colour grid  
   employed for age and metallicity
   determinations. We also study the role played by
   statistical fluctuations of the observed colours, 
   caused by the low total mass of typical globulars.}
   {Synthetic samples of globular cluster systems are created, by 
   means of Monte-Carlo techniques. Each sample accounts for a
   different possible source of bias, among the ones addressed in this investigation. 
   Cumulative age and [Fe/H] distributions are then retrieved by comparisons with a 
   reference theoretical colour-colour grid, and analyzed.
   }
   {Horizontal Branch morphology is potentially the largest source of uncertainty. 
   A single-age system harbouring a large fraction of clusters 
   with an HB morphology systematically 
   bluer than the one accounted for in the theoretical colour grid, can
   simulate a bimodal population with an age difference as large as 
   $\sim$8~Gyr. When only the redder clusters are considered, this
   uncertainty is almost negligible, unless there is an extreme mass loss
   along the Red Giant Branch phase. The metal mixture affects mainly
   the redder clusters; the effect of colour fluctuations
   becomes negligible for the redder clusters, or when the integrated 
   $M_V$ is brighter than $\sim -$8.5~mag. The treatment of 
   core convection is relevant for ages below $\sim$4~Gyr.     
   The retrieved cumulative [Fe/H] distributions are overall only 
   mildly affected. Colour
   fluctuations and convective core extension have the largest effect.
   When 1$\sigma$ photometric errors reach 0.10~mag, 
   all biases found in our analysis are erased, and  
   bimodal age populations with age differences of up to $\sim$8~Gyr 
   go undetected.
   The use of both $(U-I)-(V-K)$ and $(V-I)-(V-K)$ diagrams 
   may help disclosing the presence of blue HB stars 
   unaccounted for in the theoretical colour calibration.
}
{}

   \titlerunning{Extragalactic globular cluster ages}
   \authorrunning{M. Salaris \& S. Cassisi}
   \maketitle
%

\section{Introduction}

A large number of studies about extragalactic 
globular cluster (GC) systems has appeared in the last decades (see,
e.g. the reviews by Ashman \& Zepf~\cite{az98} and Brodie \& Strader~\cite{bs06}). 
The main motivation for these analyses is that GCs are expected to be good
tracers of major episodes of star formation of spheroids (see, e.g.,
Schweitzer~\cite{sch97}; Kissler-Patig, Forbes \& Minniti~\cite{kpfm98}) and in fact 
inferred GC age/metallicity distributions have been
widely used to investigate the formation and evolution histories of early-type
galaxies (see, e.g., Kuntschner et al.~\cite{kz02}; Larsen et
al.~\cite{lar03}; Strader \& Brodie~ \cite{sb04}; Yi et
al.~\cite{yi04}; Hempel et al.~\cite{hem05} and references therein). 
One of the most relevant discoveries in the field is the
fact that most (if not all) extragalactic GC systems show a
(integrated) colour bimodality (see, e.g., Gebhardt \& Kissler-Patig~\cite{gkp99}) 
indicating two distinct GC subpopulations, that is, two
major star-forming episodes or at least two 
different formation mechanisms in the galaxy evolution histories (see Yoon et
al.~\cite{yo06} for an alternative view on this issue).

The interpretation of this bimodality in terms of age and/or chemical
composition differences is complicated by the well known
age-metallicity degeneracy (e.g. Worthey~\cite{w94}) whereby the effect on 
the integrated colours of an age increase at fixed metallicity, can be reproduced also by
suitably increasing the metallicity while keeping the age fixed. 
The age-metallicity degeneracy can be broken in a number of ways. 
The first one is to employ spectroscopic absorption feature indices, like the Lick/IDS ones 
(Burstein et al.~\cite{bur84}; Trager et al.~\cite{trag98}). Balmer line indices are
strongly sensitive to the population age, whereas metal line indices
provide estimates of the GC chemical composition (see, e.g.,
Beasley et al.~\cite{bea00}; Strader et al.~\cite{stra05} and
reference therein for examples of application of
these indices to the study of extragalactic GCs). 
A second possibility is to use suitable broad-band colour-colour diagrams,
whereby one of the two colours is more sensitive to the age of the
population, and the other is more affected by the metallicity. A 
widely used diagram is $(V-I)-(V-K)$ (see, e.g., 
Hempel et al.~\cite{hem03}; 
Hempel \& Kissler-Patig~\cite{hem04}; Larsen, Brodie \&
Strader~ \cite{lar05}) but there are also other 
possible colour combinations, that include at least one near-IR filter
as summarized and discussed in James et al.~(\cite{jam06}) --  
see also Rejkuba~(\cite{rej01}) and Hempel et al.~(\cite{hem05}). In
addition, the $(U-B)-(B-V)$ diagram can in principle also be used 
to disentangle age and metallicity effects in case of old
populations (see, e.g., Yi et al.~\cite{yi04} for an application to the GCs 
of NGC~5128). 
A third possibility to break the age-metallicity degeneracy is to
fit the observed spectral energy distribution (SED) of the cluster --
where SED denotes in this case an ensemble of absolute magnitudes in a given
set of broad-band filters -- as discussed by  Anders et al.~(\cite{and04}) and
De Grijs et al.~(\cite{degr05}; see also Fan et al.~\cite{fan06}). In these papers the authors consider
broad-band filters ranging from $U$ to $H$, and show how a SED
composed of the $BVIH$ or $UBVI$ filters is best suited to break the
degeneracy in the GC age regime.

The most accepted view about the observed GC colour bimodality is that it is driven
mainly by metallicity (but see also Yoon et
al.~\cite{yo06} for an alternative explanation based on the morphology
of the clusters' Horizontal Branch) with the blue subpopulation being
the metal poorer one. The 
age distribution of these two subpopulations is however still a matter of
debate. As reviewed in detail by Brodie \& Strader~(\cite{bs06}) the metal
poor GCs are generally considered to be old, of age comparable to 
the age of the universe and of the oldest Galactic GCs, whereas
conflicting results are obtained for the metal rich ones.   
A representative case of this uncertainty is the GC system of
NGC~4365, recently reanalyzed in detail by Larsen et al.~(\cite{lar05}). The authors
discuss how different datasets of absorption feature indices give
contradictory results about the presence of a metal rich young GC component; also
broad-band colours provide conflicting results, due to offsets between
photometries, between theoretical integrated colours obtained
from different authors, and possible offsets 
between observed 
Galactic GC colours (especially $(V-K)$) and the theoretical counterparts. 

A definitive conclusion about the existence of a young 
component in extragalactic GC
systems is important to constrain scenarios for the
formation of spheroids. Necessary steps in this direction 
involve detailed assessments of systematic uncertainties related to the
age determination methods using absorption feature indices and
integrated magnitudes and colours. 
Here we will address methods based on integrated colour-colour
diagrams. 

A general analysis (although not
specialized to the estimate of GC ages) about uncertainties in
theoretical integrated colours of single age/single metallicity populations (also
denoted as Simple Stellar Populations -- SSPs) was published by
Charlot, Worthey \&
Bressan~(\cite{cwb96}) who compared results obtained from different sets of
stellar models and different spectral libraries. More 
recently Yi~(\cite{yi03}) discussed in a systematic way the effect of some
theoretical uncertainties on selected SSP synthetic colours.
Lamers, Anders \& de Grijs~(\cite{lam06}) have investigated 
the effect on the photometric evolution of
unresolved star clusters (for the case of solar metallicity) of 
preferential loss of low mass stars from a GC, due to mass segregation and 
evaporation. This process may alter appreciably the
integrated colours of GCs.

In this paper we focus specifically on the
$(V-I)-(V-K)$ diagram used in many studies of extragalactic GCs, like
the NGC~4365 system that -- as already mentioned -- 
provides conflicting results about the cluster age
distribution. In particular, we analyze theoretically the  
effect of a number of systematic uncertainties, on both the
inferred cumulative age distribution and the [Fe/H] cumulative
distribution. This follows the works
by, e.g., Hempel et al.~(\cite{hem03}) and Hempel \&
Kissler-Patig~(\cite{hem04}) that show how the use of the 
cumulative age distribution provides a
powerful tool to establish the presence of a young GC subpopulation. 
These authors discussed in detail the effect of the GC sample size,
observational effects due to contamination by background objects, and
the use of different population synthesis models.
To push further the analysis of this method, we investigate systematic effects 
related to the morphology of the Horizontal Branch (HB) 
that cannot be reliably predicted by theory yet; the 
metal mixture of the stars harboured by the
observed GCs (that cannot be inferred from the colour-colour diagram only);
the possible efficiency of overshooting from the convective
cores (when present) and statistical colour  
fluctuations caused by the undersampling of the bright and fast
evolutionary phases (RGB, Asymptotic Giant Branch) due to the 
low total mass of GCs. Our analysis is done differentially, with
respect to reference synthetic samples, and makes use of our
own stellar population synthesis database 
(the \textit{BaSTI} database, see Pietrinferni et al.~\cite{pietr04},
\cite{pietr06}). We want to establish, in particular, if the effects we 
examine are able to mimic a bimodal-age distribution, even in case of
a constant age for the whole cluster population. We also wish to
assess eventual biases in the cluster metallicity distribution
inferred from the colour-colour diagram, given the importance of
determining both age and metallicity distributions of GC systems 
to probe the formation histories of the parent galaxies. 

The next section presents briefly the theoretical population synthesis
models used in our analysis, and discusses the determination of integrated magnitudes and
colours. Section~3 describes the method of construction of template GC samples used as reference
for our differential study, and Section~4 presents a quantitative analysis  
of the sources of systematic uncertainties listed above. 
A summary and conclusions follow in Section~5. The Appendix provides
useful analytical relationships to determine the cumulative age
distribution of a sample of GCs from their $(V-K)-(V-I)$ colours, and a
test for the accuracy of our theoretical colour calibrations.

\section{Stellar model database and computation of integrated magnitudes/colours}

The stellar population model database adopted in our analysis is
the so-called \textit{BaSTI} (a Bag of Stellar Tracks and Isochrones) 
that can be accessed on the web at the official site:
http://www.te.astro.it/BASTI/index.php. A description of the database can
be found at the website as well as in Pietrinferni
et al.~(\cite{pietr04}, \cite{pietr06}) and Cordier et al.~(\cite{cpc06}). 
The stellar model/isochrone part of the database covers 11 different
metallicities (from extreme Pop~II stars to super metal rich
populations) and two different heavy element mixtures for each
metallicity, i.e. scaled solar and $\alpha$-enhanced typical of the 
Galactic halo population. 
We employed consistent 
bolometric corrections for both scaled-solar and $\alpha-$enhanced
mixtures, even at super-solar metallicities. 
The evolutionary tracks follow all major stages until the end of the 
Asymptotic Giant Branch (AGB) or carbon ignition,
depending on the value of the stellar mass. Mass loss from the stellar surface
is accounted for using the Reimers~(\cite{rei75}) law and two values of the
free parameter $\eta$ ($\eta$=0.2 and 0.4, respectively). 
Superwinds along the AGB phases are also
accounted for. In case of stars that 
develop convective cores during the central H-burning phase we have computed
models with and without overshoot from the Schwarzschild boundary
of the central convective regions. 

We have determined the integrated magnitudes (and colours) presented in this work as follows. 
When analytical integrated magnitudes for an SSP with a given 
age $t$ and metallicity $Z$ are needed, we
simply integrate the flux in the chosen filters along the
representative isochrone; we employ the Initial Mass Function (IMF) by Kroupa, Tout \&
Gilmore~(\cite{ktg93}) to determine the number of objects populating
each point along the isochrone, and an appropriately chosen normalization
constant. The upper mass limit is given by the most
massive star still evolving along the isochrone, the lower mass limit 
is equal to 0.5~$M_{\odot}$, that is the lower mass limit of the models in the  
\textit{BaSTI} database\footnote{We are presently working to extend the \textit{BaSTI} 
database to the regime of very low mass stars}. Objects with masses below this limit
contribute only negligibly to the integrated magnitudes and colours we
are dealing with, as we have verified by implementing the very low mass star models by  
Cassisi et al.~(\cite{cas00}), that extend down to $\sim$0.1$M_{\odot}$.
The value of the IMF normalization constant appropriate for the specified
total mass $M_t$ is obtained by enforcing the condition
$M_t=\int_{M_l}^{M_u} \psi(M) M dM$ where $\psi(M)$ is the IMF, $M$ is the stellar mass,
$M_l$ and $M_u$ the lower and upper mass values set, respectively, to
0.1$M_{\odot}$ (the contribution to the total cluster mass of objects with masses below
0.5~$M_{\odot}$ cannot be neglected)
and to the highest stellar mass still evolving in a 
population of that age and $Z$. The resulting analytical integrated colours  
are independent of the total cluster mass $M_t$.
Whenever integrated magnitudes and colours are needed for a 
($t$, $Z$) combination not available in the isochrone database, they are
determined by quadratic interpolation within our ($t$, $Z$) grid.

The analytical computation described above is somewhat the 
standard procedure followed in the literature. 
However, this procedure is strictly valid only when the number
of stars is formally infinite. Whereas the analytical
computation implies that all points along the
isochrones are smoothly populated by a number of stars that can be
equal to just a fraction of unity in case of fast
evolutionary phases, in real clusters the number of objects at a point along the Colour
Magnitude Diagram (CMD) is either zero or a multiple of unity. When the
number of stars harboured by the observed population is not
large enough to sample smoothly all evolutionary phases, statistical
fluctuations of star counts will arise. This means that an ensemble of SSPs all with the
same $M_t$, $t$ and $Z$ will show a range of integrated magnitudes
(and colours) due to stochastic variations of the number of objects populating the
faster evolutionary phases, that in case of intermediate/old SSPs 
are the upper RGB and AGB. The magnitude of these fluctuations will
be larger in the wavelength ranges most affected by the flux emitted
by RGB and AGB stars, typically the near-IR and longer wavelengths.
This is the origin of
the colour statistical fluctuations we will discuss in
the next section, and that have been addressed in the literature 
by, e.g., Chiosi, Bertelli \& Bressan~(\cite{cbb88}),
Girardi \& Bica~(\cite{gb93}), Santos \& Frogel~(\cite{sf97}), 
Cervi{\~n}o \& Valls-Gabaud~(\cite{cer03}), Cervi{\~n}o \&
Luridiana~(\cite{cl04}), Fagiolini, Raimondo \& Degl'Innocenti~(\cite{frd06}). 

To take into account the effect of these fluctuations we have computed 
integrated magnitudes and colours with a Monte-Carlo formalism. For a selected
pair ($t$, $Z$) we have first determined the appropriate isochrone by 
interpolating quadratically (when necessary) in both age and
metallicity among a set of \textit{BaSTI} isochrones. As a next step,  
star masses are randomly extracted following the Kroupa et
al.~(\cite{ktg93}) IMF, 
until a specified value of $M_t$ is reached. The position
of the individual stars along the isochrone is determined by linear interpolation
between the two neighbouring tabulated mass values (a linear
interpolation is sufficient, given the large number of points -- 2250
-- used to sample the isochrone). 
The fluxes of all synthetic stars populating the isochrone are 
then added, and magnitudes plus colours computed.

As a technical detail, we add that to compute the value of the SSP mass
consistently with the analytical procedure, the lower limit for the
random extraction of stellar masses is set to 0.1$M_{\odot}$. Whenever
a mass value between this limit and 0.5$M_{\odot}$ is extracted, it is
added to the total mass budget but disregarded for the
isochrone population.

\section{Reference GC populations}

As stated in the Introduction, the aim of this paper is to investigate the 
effect of a number of systematic uncertainties on GC age and [Fe/H] 
cumulative distributions, as inferred from the integrated $(V-I)-(V-K)$ diagram. 
As a first step we need to create some reference GC 
population that will be used for differential comparisons. We 
have therefore first created a synthetic GC system (labelled 'template' system) with the following
characteristics: 
\begin{itemize}
    \item We employed $\alpha$-enhanced isochrones (with a mean
    [$\alpha$/Fe]$\sim$0.4, the exact individual abundances of the metal
    distribution are listed in Pietrinferni et al.~\cite{pietr06}) without
    overshooting beyond the Main Sequence convective core boundaries 
    (for the appropriate mass range) and mass loss
    parameter $\eta$=0.2 constant
    for all RGB stars. The HB part of the isochrone is a clump of
    stars, because of 
    the constant value of $\eta$. This is fairly typical of
    the isochrones used in population synthesis modelling (see, e.g.,
    Girardi et al.~\cite{g00}).   
    \item The individual GC total masses ($M_t$) 
    follow the McLaughlin~(\cite{mcl94}) mass spectrum. This
    is of the form $N(M_t)\propto M_t^{-\gamma}$, 
    where we set $\gamma$ to 0.15 when the
    cluster mass is below or equal to $1.5 \times 10^5 M_{\odot}$, and
    $\gamma$=2.0 for larger masses. This is a good assumption to
    approximate the GC luminosity function (GCLF) of the Milky Way
    globulars (McLaughlin~\cite{mcl94}). We considered a mass range
    of the GC population from ${\rm log}(M_t/M_{\odot})$=3.8 up
    to 6.3.
    \item Bimodal [Fe/H] distribution. The two components are
    Gaussians centred around [Fe/H]=$-$1.55 (1$\sigma$ dispersion
    equal to 0.30~dex -- the metal poor subpopulation) and
    [Fe/H]=$-$0.55 (1$\sigma$ dispersion
    equal to 0.20~dex -- the metal rich subpopulation) respectively. 
    This distribution is an 
    approximation to the metallicity distribution of Galactic GCs
    included in the catalogue by Harris~(\cite{har96}).
    Our template sample contains 240 metal poor objects and 80 metal
    rich objects. This ratio between the two populations is
    approximately the same as in the Harris~(\cite{har96}) Galactic 
    GC catalogue. The total number of objects is taken as representative of 
    the larger extragalactic GC samples with available multicolour photometry. 
    \item Constant GC age equal to 12~Gyr. 
    \item Gaussian photometric errors with a 1$\sigma$ dispersion equal
    to 0.03~mag in all filters have been applied to the individual
    magnitudes. This uncertainty is
    comparable to the average errors in the integrated colours of
    Galactic GCs listed by Harris~(\cite{har96}), and it is taken as 
    representative of an accurate GC integrated photometry. 
\end{itemize}
The integrated colours and magnitudes of the template
population have been computed as follows. First, a cluster metallicity is
extracted randomly according to the assumed input [Fe/H] distribution, and
the age is assigned (in this case it is the same for all clusters). 
The mass $M_t$ of the cluster is then randomly selected 
according to the specified mass distribution (we do not introduce any
correlation between the cluster mass and its [Fe/H]). 
Integrated magnitudes and colours are computed analytically as
discussed in the previous
section, and then perturbed by a photometric error drawn randomly 
according to the specified Gaussian error law. 

\begin{figure}
   \centering
   \includegraphics[angle=0,width=9cm]{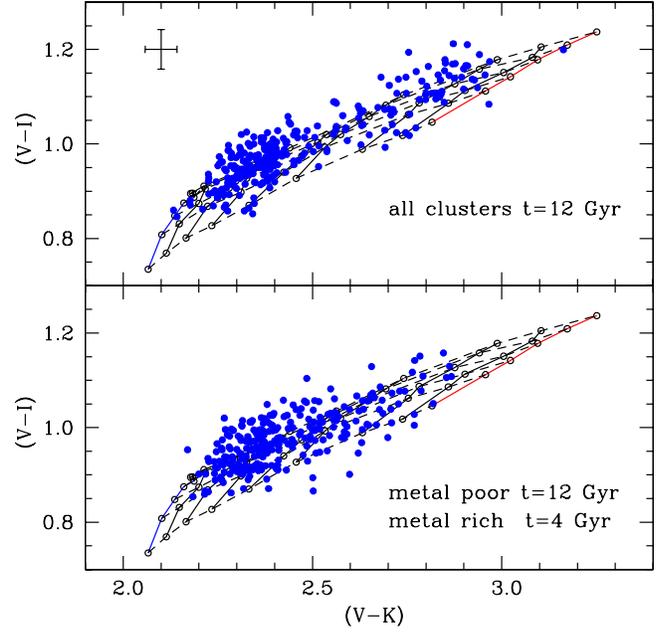}
      \caption{Integrated colours of clusters belonging to one single 
       realization of the synthetic 
       template population discussed in the text (upper
       panel). The 1$\sigma$ colour error bars are shown in the upper
   left corner of the panel. Also displayed
        is the reference grid of analytical integrated colours for 
       selected metallicities and ages. Solid lines are constant
   [Fe/H] lines, dashed lines are constant age lines. Moving towards
   increasing $(V-I)$ colours, the reference ages are equal to 2, 4, 6, 9, 12 and
   15~Gyr. Moving towards increasing $(V-K)$ colours, the 
   [Fe/H] values are equal to $-$2.62, $-$2.14, $-$1.62, $-$1.31,
   $-$1.01, $-$0.70, $-$0.29, $-$0.09, 0.05. The lower panel
   displays one single realization of the bimodal-age population A 
   discussed in the text.
              }
         \label{template}
   \end{figure}

The upper panel of Fig.~\ref{template} displays the integrated
$(V-I)-(V-K)$  (Johnson $V$- and $K$-, Cousins $I$ filter) diagram
of one single realization of the template population. Overplotted is also a
subset of the grid of analytical integrated colours used to build
the synthetic template. As already mentioned, our
analysis will focus on the age and metallicity distributions that are
obtained from the position of the clusters in this colour-colour
diagram, widely used in the literature.
To this purpose, we stress here some relevant properties of the
theoretical colour grid displayed in Fig.~\ref{template}, that will be
helpful to interpret the results presented in the rest of the
paper. As a first approximation one can assume that $(V-K)$ is 
sensitive essentially to [Fe/H], given that the $(V-I)$ range spanned by
lines of constant age and varying [Fe/H] is much smaller than the
corresponding $(V-K)$ range (see Fig.~\ref{template}). 
The $(V-I)$ colour is sensitive
mainly to age at the lowest [Fe/H] values; however, when moving 
towards the higher end of the [Fe/H] range, the lines of constant
[Fe/H] and varying age span also a sizable range of $(V-K)$ colours.  

The theoretical grid covers a $(V-K)$ interval 
$\sim$2-3 times 
larger than the corresponding $(V-I)$ range. This implies that
similar changes of $(V-K)$ and $(V-I)$ colours will affect
comparatively more the age determination 
(largely sensitive to $(V-I)$) than the [Fe/H] estimates. 
It is also important to notice that in the low [Fe/H] regime the lines 
corresponding to constant ages above 9~Gyr lie closer 
than at higher [Fe/H]. This implies that small changes of $(V-I)$  
cause larger variations of the retrieved age of old populations, 
compared to the case of higher [Fe/H].  
%

Coming back to Fig.~\ref{template}, 
an obvious point is to notice the important effect of the photometric
error, even with a 1$\sigma$ error as small as 0.03~mag. Some objects  
appear younger than 2~Gyr when dated according to their location 
in the diagram, and many clusters appear older than
15~Gyr. Also the individual metallicities retrieved are obviously  
affected by the photometric errors, but the effect on the age is the 
most striking. 
This shows clearly that age-dating only a handful of objects in an
extragalactic GC population can lead to erroneous results about the age
distribution of the cluster system. Even when larger samples are
observed, it is paramount that their colour-colour distribution is not
biased compared to the case for the whole system.

As a second reference sample used in the comparisons that follow 
in the next section, we have created a synthetic GC system with an intrinsic 
bimodal-age population (labelled 'population A'). This system has the same properties of 
the template sample, the only difference being that the metal rich
component is 4~Gyr instead of 12~Gyr old. The integrated colour-colour diagram of 
one realization of this bimodal-age population is displayed in the
lower panel of Fig.~\ref{template}; 
the analytical colour grid used to create this sample A is 
obviously the same employed for the template population, and it is also 
plotted in the diagram.
To assess quantitatively the differences in the age distributions retrieved from the
colour-colour diagram, we employ the cumulative age distributions -- introduced by, e.g., Hempel
et al.~(\cite{hem03}, \cite{hem04}) to establish statistically the
existence of young subpopulations -- as obtained from the individual cluster
colours and the reference grid displayed in Fig.~\ref{template}.  
The cumulative age
distribution (CAD -- number fraction of clusters with age larger than $t_i$
as a function of $t_i$) for one single realization of a synthetic cluster
populations has been computed as follows. 
We associate to each cluster an age greater than $t_i$ when
it lies above the line of constant $t_i$ (we considered $t_i$ values
equal to 0, 2, 4, 6, 9, 12 and 15~Gyr) in the $(V-I)-(V-K)$
diagram. Objects below the 2~Gyr line are assigned to the bin
corresponding to $t>t_i$=0~Gyr. 
Given the $(V-K)$ of an individual synthetic GC, we 
determine by linear interpolation within the reference grid the expected
$(V-I)$ of the two neighbouring
constant age lines at the cluster $(V-K)$. 
After this, it is straightforward to use the cluster $(V-I)$  
to assign the object to the appropriate CAD age intervals. 
For the very few synthetic clusters -- number of the order of unity -- that appear in
some realization beyond the red and blue $(V-K)$ edges of the theoretical
calibration, we extrapolate the constant age reference colours by fitting
the last three points of a given $t_i$ sequence 
with a function of the form $(V-I)_{t_i}=a + b \ {\rm ln}[(V-K)]$. 
The individual CAD number counts are then normalized to
the total number of objects in each realization.
In the Appendix we provide accurate fitting formulae $(V-I)_{t_i}=f[(V-K)]$
that can be applied to the whole $(V-K)$ range spanned by our
theoretical colours, and allow a fast but still reliable estimate of
the age distribution from the measured $(V-K)$
colours, without the need to interpolate within the reference colour grid.

\begin{figure}
   \centering
   \includegraphics[angle=0,width=8cm]{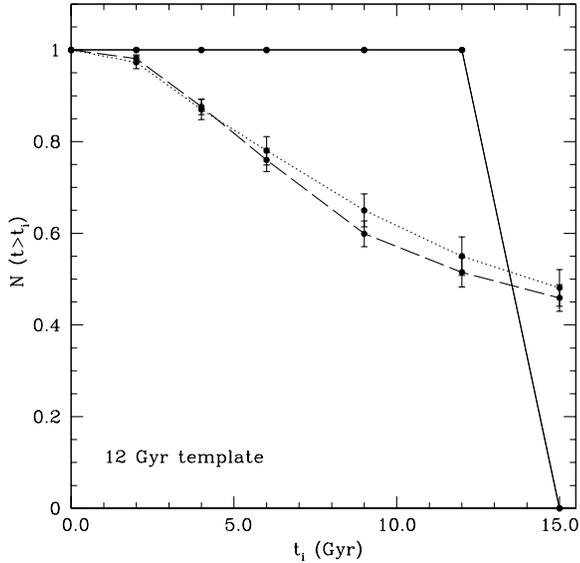}
      \caption{CADs for the 12~Gyr template sample. The solid line
       displays the real age distribution, dotted and dashed lines the
   CAD retrieved from the colour-colour diagram for the full sample
   (dashed line) and for only the clusters with $(V-K)>$2.4 (dotted
   line -- see text for details).  
              }
         \label{clf1}
   \end{figure}

\begin{figure}
   \centering
   \includegraphics[angle=0,width=9cm]{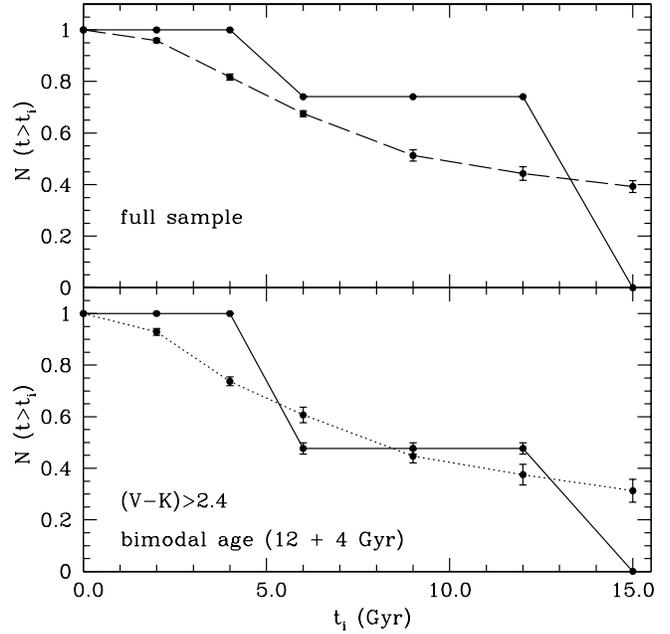}
      \caption{As in Fig.~\ref{clf1} but for population A with
   the bimodal age distribution. The upper panel shows the CADs for the
   whole sample (real age distribution -- solid line; retrieved age
   distribution -- dashed line) whereas the lower panel shows the age
   distribution only for clusters with $(V-K)>$2.4 (real age
   distribution -- solid line; retrieved age distribution -- dashed line).  
              }
         \label{clf2}
   \end{figure}

Figures~\ref{clf1} and \ref{clf2} show the CADs obtained from 
30 realizations of, respectively, the template and the bimodal-age
samples discussed above. Filled circles represent the mean of the 
results obtained from the multiple realizations, error bars show
the 1$\sigma$ dispersion around these mean values.
The size of the error bars associated to the individual points of the CAD 
is affected -- keeping everything else unchanged -- by the GC sample size, 
and increases when the number of synthetic GCs decreases   
(see also Hempel \& Kissler-Patig~\cite{hem04}).

%
%

Figure~\ref{clf1} shows very clearly the large change in the age
distribution retrieved from the colour-colour diagram, compared to the
input one (solid line). The ages retrieved for the whole sample (dashed line) 
span a range between 2~Gyr (or less) up to
more than 15~Gyr, and the shape of the CAD is greatly changed,
displaying a smooth decrease with $t_i$, because the synthetic clusters tend to be
more evenly distributed over the reference clour grid. 
If we restrict the sample to the more metal
rich objects, with $(V-K)>2.4$ (about half of the total sample  
falls into this colour interval, that corresponds to retrieved [Fe/H] 
values larger than 
$\sim -$ 1.0 to $-$1.3, using the calibration provided by the reference grid) 
the CAD is very similar, although not exactly  
identical. This is due to the different sensitivity of the
colours to age (and also to metallicity) at different
locations on the diagram. 
The case for the bimodal age distribution  is displayed in
Fig.~\ref{clf2}. Again, the retrieved ages show a CAD very
different from the real one; the trend with $t_i$ is, as expected,
much smoother. Restricting the analysis the sample to objects redder
than $(V-K)>2.4$ allows one to study a sample with a
higher ratio of old to young objects, because we are mainly sampling
the higher metallicity and younger clusters in the synthetic 
populations\footnote{The 1$\sigma$
dispersion of the input age distribution in the age range between 6 and 12~Gyr is
due to the fluctuation of the number of objects populating the
overlapping tails of the Gaussian metallicity distributions, and also
to the effect of the photometric error that moves objects belonging to the
two subpopulations in and out of this colour range}.
   \begin{figure}
   \centering
   \includegraphics[angle=0,width=9cm]{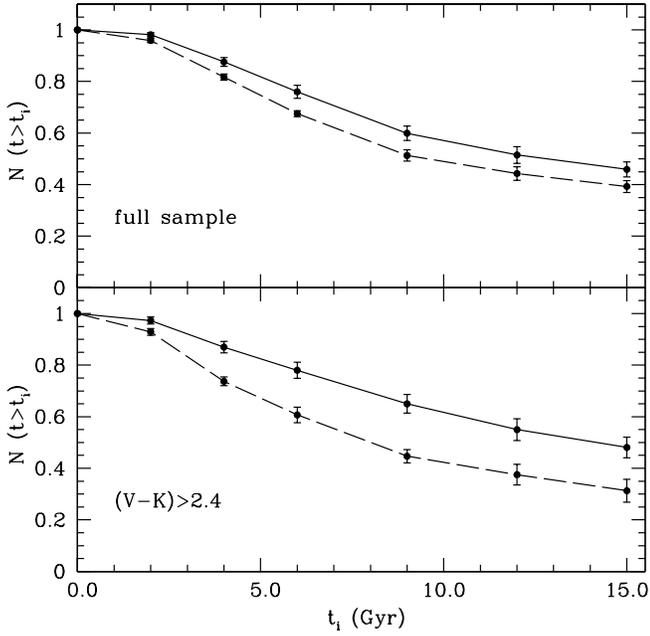}
      \caption{Comparison between the retrieved CADs for the 12~Gyr template
    (solid line) and the bimodal-age population A (dashed line). The
     upper panel displays the comparison for the full synthetic samples, the
    lower panel the comparison only for objects with $(V-K)>$2.4. 
              }
         \label{clf3}
   \end{figure}
%

Figure~\ref{clf3} displays a comparison between the retrieved CADs of the two 
populations, for the full samples (upper panel) and for
objects with $(V-K)>$2.4 only (lower panel). The
difference between the two CADs is evident, in both the full and
restricted samples. Bimodal-age CADs are always below the unimodal ones. 
This is the basis for the use of CDAs as indicators of age bimodality,
as discussed in detail by Hempel \& Kissler-Patig~(\cite{hem04}). 
The sample with $(V-K)>$2.4 shows a larger difference between the two
CADs, because in this colour range the ratio young/old
clusters in the bimodal-age population is about 50\%, 
whereas when the full sample is considered, it is only 25\%.

The comparison presented in Fig.~\ref{clf3} summarizes the main results  
of this discussion about the CAD of our two reference 
synthetic populations. Bimodal-age (young+old) populations display a 
CAD systematically lower than the case of unimodal (old) populations. The extent of 
this difference depends on both the age of the 
young subpopulation and the number ratio between the two components. Increasing the 
ratio of young to old clusters -- keeping the age distribution fixed -- shifts the CAD 
towards increasingly lower values. The same happens when one decreases 
the age of the young component -- keeping constant the number ratio between the two subpopulations -- 
as we will show explicitely during one of the tests of the next section  
(analogous conclusions can be found in Hempel \& Kissler-Patig~\cite{hem04}).

\begin{figure}
   \centering
   \includegraphics[angle=0,width=9cm]{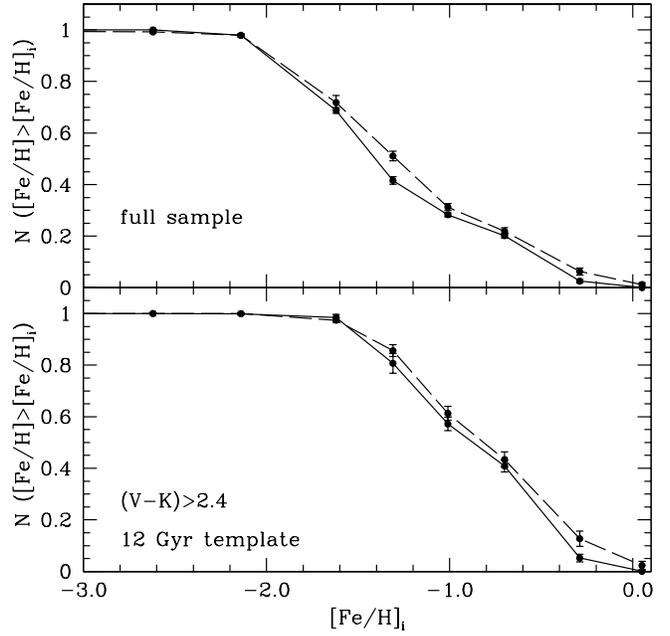}
      \caption{CMeDs for the 12~Gyr template sample. The solid line
       displays the real [Fe/H] distribution, dashed lines the
       CMeD retrieved from the colour-colour diagram for the full sample
       and for only the clusters with $(V-K)>$2.4. 
       Here, as in all following figures displaying CMeDs, we
       do not show the bin corresponding to ${\rm [Fe/H]}_i> -4.5$, that
       always has a normalized number count equal to 1.  
              }
         \label{mlf1}
   \end{figure}

\begin{figure}
   \centering
   \includegraphics[angle=0,width=9cm]{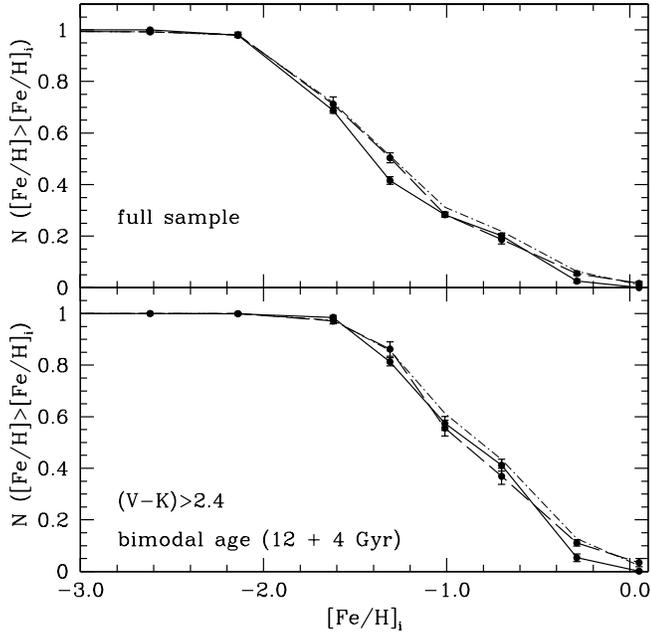}
      \caption{As in Fig.~\ref{mlf1} but for population A with
   a bimodal age distribution. For comparisons we also
   display -- dashed-dotted line -- the corresponding CMeDs retrieved in
   the case of the 12~Gyr template. 
              }
         \label{mlf2}
   \end{figure}

Figures~\ref{mlf1} and ~\ref{mlf2} display the cumulative [Fe/H]
distribution (CMeD) from the same multiple realizations of the
template and the bimodal-age A sample.
As in case of the CAD, filled circles represent
the mean of the results obtained from the various realizations, 
and error bars show the 1$\sigma$ dispersion around these mean values.
For each realization we associate to an individual cluster an ${\rm
[Fe/H]}>{\rm[Fe/H]}_i$ when it lies to the right of the line of
constant ${\rm [Fe/H]}_i$ predicted by the reference grid. We considered 
${\rm [Fe/H]}_i$ values equal to $-$2.62, $-$2.14, $-$1.62, $-$1.31,
$-$1.01, $-$0.70, $-$0.29, $-$0.09, 0.05, as in Fig.~\ref{template}.
As for objects blueward of the [Fe/H]=$-$2.62 line -- number usually
of the order of unity -- they are included into an additional bin corresponding to 
[Fe/H]$> {\rm [Fe/H]}_i=-4.5$, that is an arbitrary low value. The individual
number counts of each CMeD are then normalized to
the total number of objects in each realization.
Similar to the case of the CAD, we determine the appropriate [Fe/H] bin of the
individual objects by linear interpolation among the reference
colour grid. For synthetic objects above the highest age of the
reference grid, we extrapolate the constant [Fe/H] lines by
approximating $(V-K)$ as a linear function of $(V-I)$, fitted to the last three
points below the grid boundary, for each
reference metallicity. For objects located below the 2~Gyr line we supplement the grid
shown in Fig.~\ref{template} with the integrated
colours of a 1~Gyr old population. If necessary, slightly younger ages
are accounted for. 

%
%

The CMeDs plotted in Figs.~\ref{mlf1} and ~\ref{mlf2} demonstrate how the input
and retrieved CMeDs show only marginal differences. In the case of both the
full sample and the $(V-K)>$ 2.4 one, input and retrieved
metallicity distributions are very similar for both age
distributions. The largest difference is
at ${\rm [Fe/H]}_i=-$1.31 for the full samples, where the retrieved
CMeDs show counts larger then the intrinsic one. This is due to the
fact that the intrinsic [Fe/H] 
distribution shows a dip between the peaks associated to the two
metallicity components, that falls into this [Fe/H] bin; the
retrieved [Fe/H] values have however a much 
less pronounced dip due to the smoothing effect of the photometric
error. 
An important result is that the different age distribution 
has only a small effect on the 
derived CMeD. This is clearly illustrated by Fig.~\ref{mlf2}, 
where the CMeDs retrieved from the template sample are also displayed.

\section{Analysis of systematic effects}

In this section we analyze to what extent systematic
effects related to the unknown morphology of the horizontal branch in
unresolved GCs, their unknown metal distribution, colour statistical
fluctuations and the uncertain efficiency of convective core
overshooting, are able to change the CAD retrieved using a fixed reference
grid of analytical integrated colours. 
As already mentioned in the introduction, our main goal is to establish
whether these effects are able to mimic a bimodal-age distribution in
presence of an uniformly old stellar population. 
We also wish to assess the possible occurrence 
of biases in the cluster metallicity distribution
inferred from the colour-colour diagram, as compared to the intrinsic one.

The general method of analysis is the following. We create synthetic
populations with the same $t$, [Fe/H], $M_t$, photometric error
distributions of a reference system -- 
this latter determined using our reference colour grid displayed in Fig.~\ref{template}; a
reference system could be our template 
single-age or the bimodal-age population A described in the
previous section -- but using integrated magnitudes that include the effect
of, for example, a different HB morphology. 
Table~\ref{table1} summarizes the main properties of all the synthetic GC samples  
employed in these tests.

We then retrieve the age and [Fe/H] distributions of these new  
synthetic samples by employing our reference grid 
(that  does not take into account, e.g., the different HB
morphology) and compare the results with those obtained for the
reference systems.

%
\begin{table*}
\caption{Main properties of the synthetic GC samples discussed in this paper (see text for details). }             
\label{table1}      
\centering          
\begin{tabular}{c l l l l l l l}     
\hline\hline       
Label & Mass + [Fe/H]  & Age & Mass loss & Convection & Metal mixture & 1$\sigma$ phot. error & Colour computation\\ 
\hline                    
Template & Milky Way GCs& 12~Gyr   & $\eta$=0.2            &No-oversh. &$\alpha$-enh.                &0.03 mag & Analytic   \\  
   A     & Milky Way GCs& 12+4~Gyr & $\eta$=0.2            &No-oversh. &$\alpha$-enh.                &0.03 mag & Analytic   \\
   B     & Milky Way GCs& 12~Gyr   & $\eta$=0.4            &No-oversh. &$\alpha$-enh.                &0.03 mag & Analytic   \\
   B2    & Milky Way GCs& 12~Gyr   &bimodal $\eta$=0.2,0.4 &No-oversh. &$\alpha$-enh.                &0.03 mag & Analytic   \\
   C     & Milky Way GCs& 12~Gyr   & $0.2\le \eta \le$ 0.4 &No-oversh. &$\alpha$-enh.                &0.03 mag & Analytic   \\
   D     & Milky Way GCs& 12~Gyr   & $\eta$=0.2            &No-oversh. &$\alpha$-enh. + scaled solar &0.03 mag & Analytic   \\
   E     & Milky Way GCs& 12~Gyr   & $\eta$=0.2            &No-oversh. &$\alpha$-enh.                &0.03 mag & Monte-Carlo\\  
   F     & Milky Way GCs& 12+4~Gyr & $\eta$=0.2            &No-oversh. &$\alpha$-enh.                &0.03 mag & Monte-Carlo\\
   G     & Milky Way GCs& 12+2~Gyr & $\eta$=0.2            &No-oversh. &$\alpha$-enh.                &0.03 mag & Analytic   \\
   H     & Milky Way GCs& 12+2~Gyr & $\eta$=0.2            &Oversh.    &$\alpha$-enh.                &0.03 mag & Analytic   \\
   I     & Milky Way GCs& 12~Gyr   & $\eta$=0.2            &No-oversh. &$\alpha$-enh.                &0.10 mag & Analytic   \\
   J     & Milky Way GCs& 12~Gyr   & $\eta$=0.4            &No-oversh. &$\alpha$-enh.                &0.10 mag & Analytic   \\  
\hline                  
\end{tabular}
\end{table*}

The quantitative results of these comparisons are inevitably dependent on the
assumed photometric error, [Fe/H], age and -- in some case -- $M_t$ distribution of the
synthetic GCs. 
With the exception of the photometric error, that when 
largely increased tends to wash out all other effects, the general
trends should be unaffected by  
the precise choices made about these parameters, and provide useful
guidelines to understand the impact of these uncertainties on our
interpretation of extragalactic GC systems. 
We employed throughout our analysis the [Fe/H], $t$ and $M_t$
distributions of a real GC system,
e.g. the Galactic one, already introduced in the previous section. 
As for the photometric error, we continue to use a 
1$\sigma$=0.03~mag value, that
corresponds to accurate photometry of Galactic GCs. The size of our
synthetic samples are also large -- the same as the template and the
bimodal-age ones introduced in the previous section. 
This combination of -- realistic -- 'small' photometric error and
large sample size, highlights more clearly the role played by the 
uncertainties we are going to address.  
As mentioned previously, the 
effect of reducing the sample size is purely to increase the error bars 
attached to the individual points of the retrieved CADs and CMeDs. 
The effect of increasing photometric
errors will be addressed at the end of this section.

\subsection{Horizontal Branch morphology}
\label{eta}

To date, the theory of stellar evolution is unable to predict from first
principles the expected morphology of the HB in an old stellar
population of fixed age and chemical composition. This arises from the lack of a
theory for the stellar mass loss along the RGB phase. Once age and
chemical composition are fixed, the HB morphology is completely determined by
the amount of mass lost along the RGB phase by the progenitor
objects. From synthetic HB modelling we know that stars along the
RGB phase have to lose on average $\sim$0.10--0.20$M_{\odot}$ in order to 
reproduce the typical colour location of
the HBs observed in Galactic GCs, and a 1$\sigma$ spread of the order of
$\approx$0.02$M_{\odot}$ around this mean mass
loss value has to be typically considered to match the observed HB colour
extension (see, e.g., Lee, Demarque \& Zinn~\cite{ldz94}). However,
the precise values may differ from cluster to cluster and are in
principle unknown for extragalactic unresolved GCs.  

Observational determinations of mass loss rates along the
RGB are scarce (see, e.g. Origlia et al.~\cite{of02}) and have not led
yet to a mass loss theory with predicting power. The
Reimers~(\cite{rei75}) mass loss formula is traditionally used 
in stellar evolution modelling; it contains a free parameter $\eta$,
whose value determines the efficiency of the mass loss process (see
also Catelan~\cite{cat06} for a summary of alternative, but less used
mass loss formulae).  

   \begin{figure}
   \centering
   \includegraphics[angle=0,width=9cm]{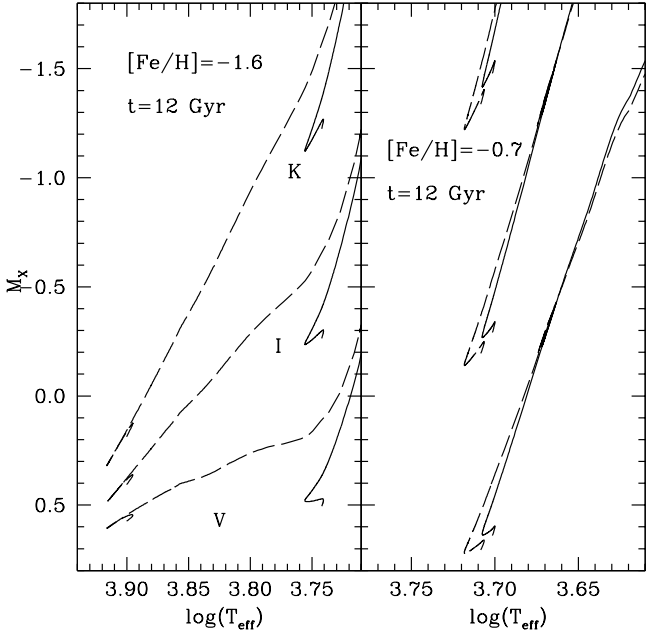}
      \caption{Comparison of the HB evolution in the
   magnitude-effective temperature plane, for 12~Gyr old isochrones 
   computed with the mass loss parameter $\eta$=0.2 (solid line) 
   and 0.4 (dashed line) respectively, and the two labelled metallicities.
   The $V$, $I$ and $K$ labels denote the photometric filter used 
  for each of the three pairs of tracks represented in each of the 
  two panels.
  }
         \label{hb1}
   \end{figure}
%

A quantitative example of the effect of different
RGB mass loss assumptions on the HB magnitudes and colours is shown by  
Figure~\ref{hb1}. Here we display the
magnitude-$T_{eff}$ HB sequence for a 12~Gyr
$\alpha$-enhanced isochrone and the two labelled [Fe/H] values (that
correspond approximately to the mean values for the two metallicity
components of our synthetic GC populations) for 
$\eta=0.2$ (the value adopted in our template GC sample) and 0.4.
These two values of $\eta$ produce a total RGB mass loss equal to,
respectively, 0.08$M_{\odot}$ and 0.17$M_{\odot}$ at [Fe/H]=$-$1.6; 
0.10$M_{\odot}$ and 0.20$M_{\odot}$ at [Fe/H]=$-$0.7.
Larger mass loss implies a hotter HB, due to the smaller ratio 
between total mass and He-core mass; this can be clearly seen in the figure, together
with the fact that the 
effect is much larger at the lower metallicity. The magnitude of the
Zero Age HB (ZAHB) location (stars spend most of their 
HB evolution near the ZAHB location) becomes fainter by a 
large amount in $I$ (about 0.8~mag) and even more in $K$ (about 1.5~mag) when $\eta$
is doubled and [Fe/H]=$-$1.6. As a general rule 
and for all metallicities, the size of the magnitude change depends on the photometric
filter one considers.

Uncertain predictions of the HB morphology in unresolved GCs lead to
uncertainties in the predicted integrated magnitudes and colours, given that
the HB phase is bright and relatively long-lived, and accounts for a
non-negligible fraction of the cluster integrated flux. 
   \begin{figure}
   \centering
   \includegraphics[angle=0,width=9cm]{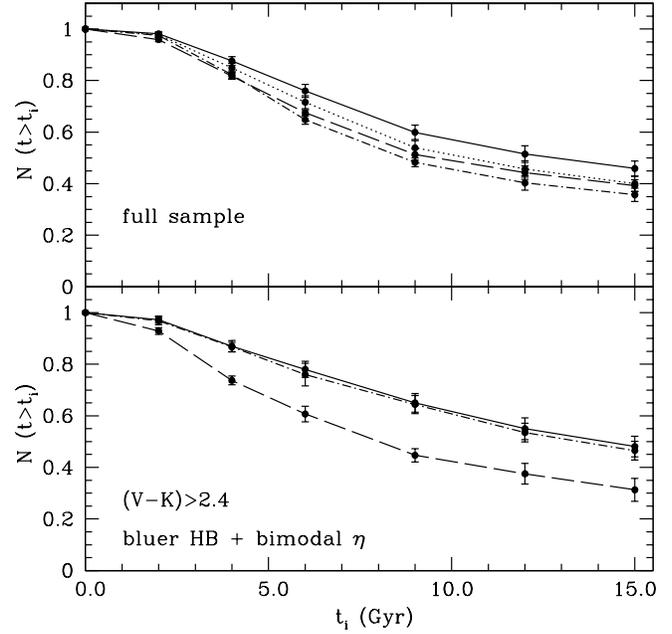}
      \caption{Comparison between the retrieved CADs for the 12~Gyr template
    (solid line) the $\eta=0.4$ population B (dashed-dotted lines) and the bimodal $\eta$ population B2 
    (dotted line). The
    upper panel displays the comparison for the full synthetic samples, the
    lower panel the comparison only for objects with $(V-K)>$2.4 (the results for population 
    B2 are not displayed in this panel, because they are identical to the case of the template population). 
    The corresponding CADs for the bimodal-age population are also
    displayed as dashed lines.
              }
         \label{clf4}
   \end{figure}
%

To give a quantitative estimate of how a wrong assumption about the
unknown HB colour affects the ages derived from the $(V-K)-(V-I)$
plane, we first created a
synthetic GC population labelled 'population B') with exactly the same properties of the 
12~Gyr template one, but employing isochrones computed with $\eta$=0.4,
i.e., with a bluer HB. We then determined the CAD for 30 realizations
of this sample, using
the reference grid of our template population (with $\eta$=0.2). 
The resulting CAD is displayed in Fig.~\ref{clf4},
compared to the results for the
template population and the bimodal-age population A discussed in
the previous section.

When the whole cluster samples are compared, population B 
shows an age distribution very similar to the bimodal-age population A, 
whereas if we consider only objects with
$(V-K)>$2.4, there is essentially no difference compared to the result
for the template population. This is because at [Fe/H]$<
-$1.6 the bluer HB morphology affects strongly the integrated colours,
that become bluer and mimic a much younger age if the CAD is determined from the
reference ($\eta$=0.2) colour grid. This effect is negligible at higher
metallicities because the HB colour and magnitude change only slightly when doubling
$\eta$ (see Fig.~\ref{hb1}). Of course, if the mass loss is so high as to produce
a much bluer HB also at [Fe/H]$> -$1.0, then we would retrieve
substantially young ages also for the cluster with $(V-K)>$2.4.  

   \begin{figure}
   \centering
   \includegraphics[angle=0,width=9cm]{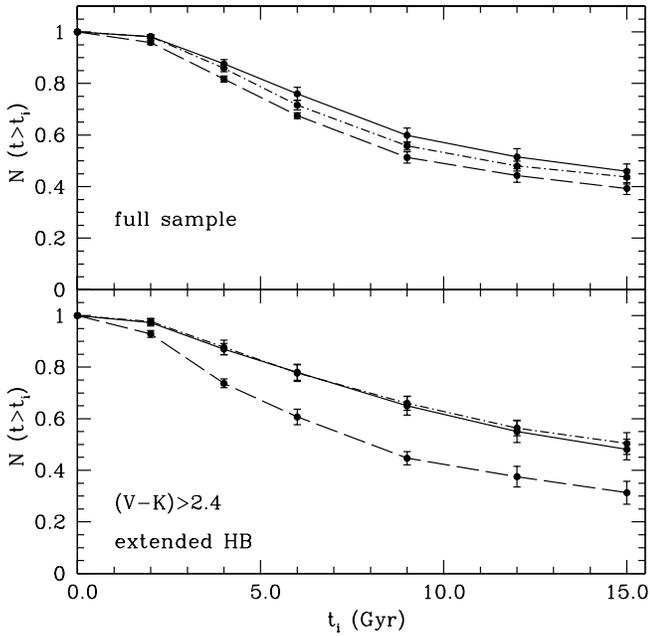}
      \caption{As in Fig.~\ref{clf4}, but for the comparison with the 
       12~Gyr old, extended HB population C (dashed-dotted line).
              }
         \label{clf5}
   \end{figure}
%
As a second experiment, we tested the influence of having clusters with the same [Fe/H] but 
different mass loss efficiencies along the RGB. This effectively means that GCs with the same 
metallicity (and age) can have different HB morphologies (as actually observed in the Galaxy, like in case of the 
cluster pair M3-M13). To this purpose we created a 12~Gyr 
GC population (labelled 'population B2') with the same properties of the template sample, 
but where 50\% of the synthetic clusters with [Fe/H]$< -$1.1 (we saw before that changing $\eta$ from 0.2 to 0.4 
does not have an appreciable effect on the retrieved CAD at higher metallicities) is originated 
from isochrones computed with $\eta$=0.4, and the remaining 50\% from isochrones with with $\eta$=0.2 
(bimodal $\eta$ distribution). The retrieved CAD is displayed in the upper panel of Fig.~\ref{clf4} 
for 30 realizations of this sample, using
the reference grid of our template population (with $\eta$=0.2). 
The CAD lies, as expected, 
above the case of population B ($\eta$=0.4 for all clusters) but it is still very close 
to the bimodal-age population A for ages above 9~Gyr. Of course, a reduced fraction of clusters with 
$\eta$=0.4 would move the CAD towards the template case.

As a third quantitative test, we have determined a synthetic population with an HB
extended in colour (labelled 'population C') more similar to the HBs observed in real GCs. 
The mass distribution along the HB is assumed to have a flat profile 
between the values corresponding to $\eta$=0.2 and
0.4. Figure~\ref{clf5} compares the retrieved CADs with the template
and the bimodal-age population A. Predictably, the effect on the whole sample
is smaller, and the CADs is halfway between the template 
and population A. Clusters with $(V-K)>$2.4 are unaffected.
A very similar CAD, slightly closer to the template case, is found from a synthetic GC sample 
identical to population C, but with an HB morphology obtained employing   
a mean HB mass obtained from a $\eta$=0.3 mass loss parameter along the RGB, and 
a 1$\sigma$ dispersion of 0.015$M_{\odot}$ around this mean value
\footnote{This additional test has been performed 
making use of the extended HB model library included in the \textit{BaSTI} database}.  
This type of mass distribution produces a relationship between [Fe/H] and the HB morphology parameter 
$HB_t=(B-R)/(B+V+R)$ (where $B$, $V$ and $R$ are in this case the number of 
HB stars bluer than the RR~Lyrae instability strip, 
within the strip, and redder than the strip, respectively. We employed the relationships by 
Di Criscienzo, Marconi \& Caputo~\cite{dic04} for identifying the position of the RR~Lyrae instability 
strip in our simulations) 
close to the mean relationship for Galactic GCs, 
as obtained from data in the Harris~(\cite{har96}) catalogue. 

The main conclusion of all these tests is that the unknown morphology of the
HB in an unresolved GC population has the potential to strongly affect 
the cluster CADs -- particularly at low-intermediate 
metallicities -- and mimic an old+young bimodal-age population, even
in the case where all clusters are old.

   \begin{figure}
   \centering
   \includegraphics[angle=0,width=9cm]{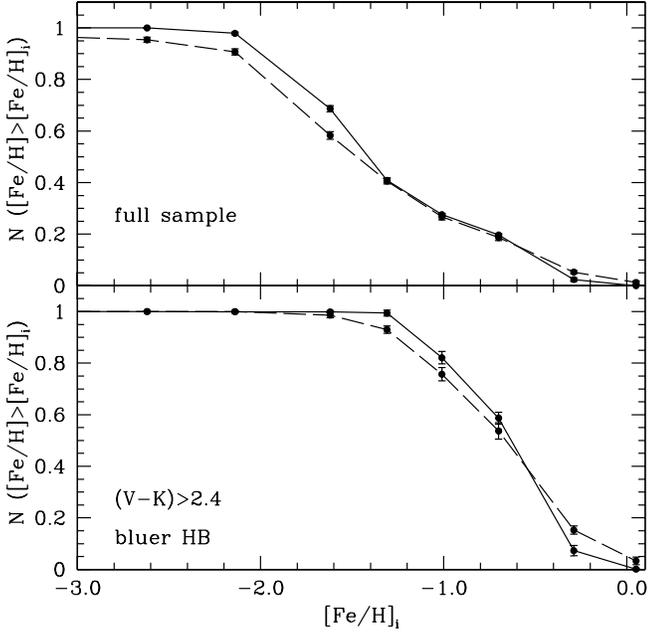}
      \caption{As in Fig.~\ref{mlf1} but for the comparison with the 
   12~Gyr old population B with
   $\eta$=0.4 mass loss law (see text for details). 
              }
         \label{mlf3}
   \end{figure}
%

   \begin{figure}
   \centering
   \includegraphics[angle=0,width=9cm]{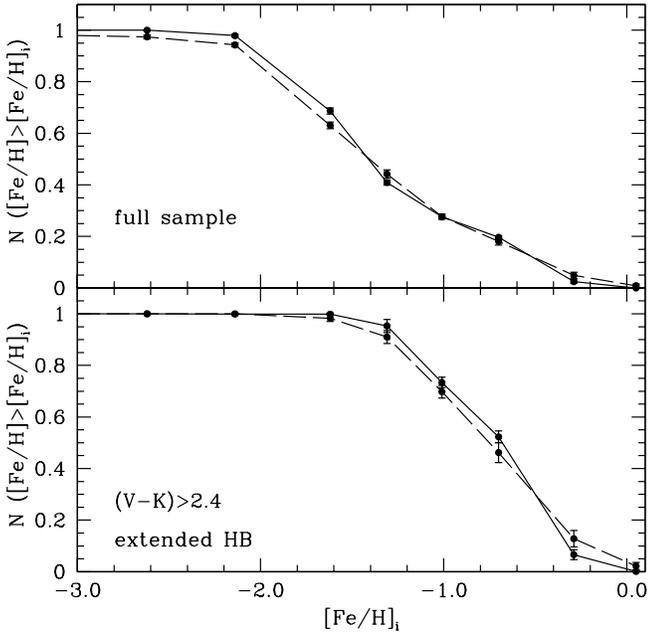}
      \caption{As in Fig.~\ref{mlf1} but for the 12~Gyr old population C with
   the extended HB (see text for details). 
              }
         \label{mlf4}
   \end{figure}
%

The retrieved CMeDs for the two blue-HB synthetic populations B and C  
are displayed in Figs.~\ref{mlf3} and
\ref{mlf4}, compared to the intrinsic ones. 
Differences with respect to the intrinsic CMeDs are generally small. 
At lower [Fe/H] below $\sim -$0.7 the retrieved CMeDs
display systematically (slightly) lower number fractions compared 
to the intrinsic values, and it is
due to the effect of a larger mass loss, that affects mainly the
low-intermediate metallicities, and causes bluer integrated colours. This affects
not only the age estimate, but also the retrieved [Fe/H] values.
The CMeD for the sample B2 is intermediate between the results 
for populations B and C.

\subsection{Scaled solar vs $\alpha$-enhanced mixtures}

The initial metal distribution of extragalactic GCs for which we are able to
measure only integrated colours is in principle unknown. One can use as
a guideline the fact that the Galactic GC system displays an
$\alpha$-enhanced metal mixture (see, e.g., Carney~\cite{c96}) but
this might not be the case in other GC systems\footnote{We notice that theoretical
calibrations of colour-colour diagrams employed to determine
GC ages are usually based on scaled solar isochrones}.  

   \begin{figure}
   \centering
   \includegraphics[angle=0,width=9cm]{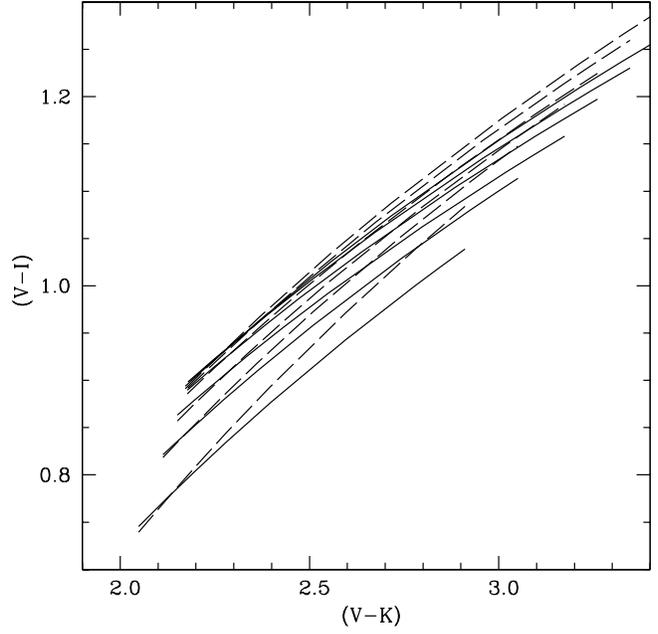}
      \caption{Lines of constant age in the $(V-K)-(V-I)$ diagram for 
    scaled solar (solid lines) and $\alpha$-enhanced (dashed lines) 
    metal mixtures, as obtained from the analytical relationships
   given in the Appendix. 
   Ages are the same as in Fig.~\ref{template} and increase from
   bottom to top. 
              }
         \label{grid}
   \end{figure}
%

We display in Fig.~\ref{grid} lines of constant age in the
$(V-K)-(V-I)$ diagram, for both scaled solar and $\alpha$-enhanced
metal mixtures, in the $(V-K)$ range spanned by the $\alpha$-enhanced 
calibration. For ease of interpretation, 
the smooth constant-age lines have been determined using 
the analytical relationships given in the Appendix. The scaled solar
colours have also been obtained from models in the \textit{BaSTI}
database and are homogeneous with the $\alpha$-enhanced ones. 
The differences in the two theoretical calibrations are evident. 
At the bluest end (i.e. at the lowest [Fe/H]) of the theoretical calibration the 
colour differences are small, but they increase sharply with
increasing $(V-K)$, e.g, [Fe/H]. The net effect is to assign too young
ages to GCs with scaled solar metal mixtures when an
$\alpha$-enhanced calibration is used. Of course the reverse is true
when determining ages of $\alpha$-enhanced GCs with a scaled solar
calibration. 

   \begin{figure}
   \centering
   \includegraphics[angle=0,width=9cm]{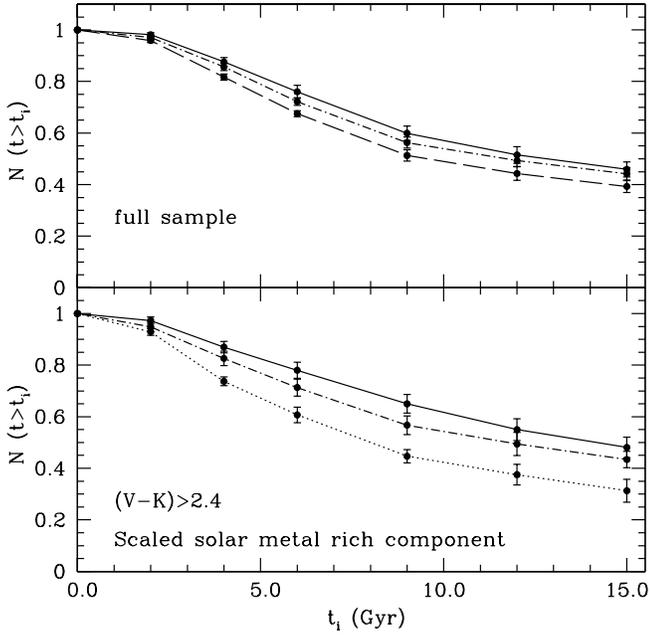}
      \caption{Comparison between the retrieved CADs for the 12~Gyr template
    (solid line) and population D (dashed-dotted lines) whose metal rich subpopulation is assumed to be 
   born out of a scaled solar metal distribution. The
    upper panel displays the comparison for the full synthetic samples, the
    lower panel the comparison only for objects with $(V-K)>$2.4. 
    The corresponding CADs for the bimodal-age population A are also
    displayed as dashed lines.
              }
         \label{clf6}
   \end{figure}
%

To give more quantitative estimates of this effect we have created a
synthetic sample 
(labelled 'population D') with the same properties as the 12~Gyr template one, but employing a scaled
solar metal mixture for the metal rich component, i.e. the component
centred around [Fe/H]=$-$0.55. We then retrieved
the CADs and CMeDs of this system with our reference $\alpha$-enhanced colour grid. 
Figure~\ref{clf6} compares the CADs for this whole new sample and the 
$(V-K)>$2.4 clusters only, with the corresponding template and
bimodal-age population A ones. The ``erroneous'' metal mixture attributed to the
metal rich component simulates a bimodal-age population, albeit with a smaller age
difference than our 12+4 Gyr system. The effect is clearly enhanced when the sample is
restricted to only the more metal rich clusters with $(V-K)>$2.4.   

   \begin{figure}
   \centering
   \includegraphics[angle=0,width=9cm]{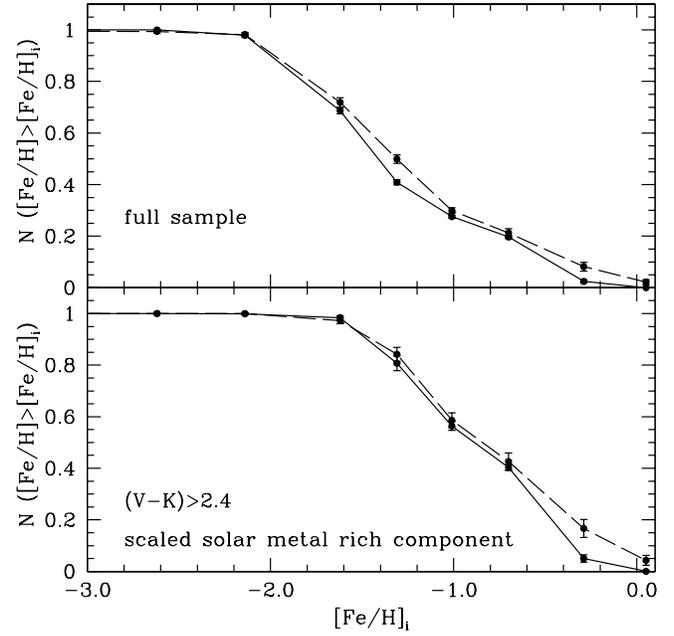}
      \caption{As in Fig.~\ref{mlf1} but for the 12~Gyr old Population D 
   with a scaled solar metal rich subpopulation. 
              }
         \label{mlf5}
   \end{figure}
%

The corresponding CMeDs are compared in Figure~\ref{mlf5}. The
differences with respect to the intrinsic distributions are small,
almost identical to the case of the template and bimodal-age samples
shown in Figs.~\ref{mlf1} and \ref{mlf2}. This means that the metal
mixture does not influence appreciably the retrieved [Fe/H]
distributions, at least for the two specific mixtures used in this test.

   \begin{figure}
   \centering
   \includegraphics[angle=0,width=9cm]{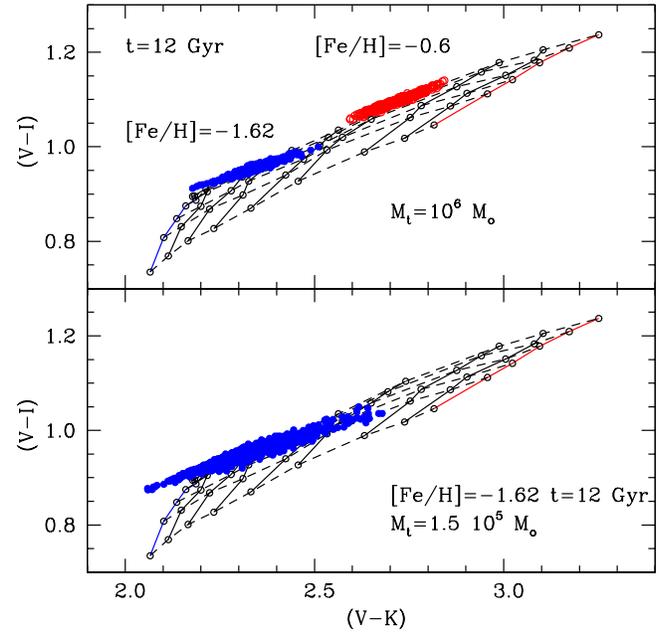}
      \caption{Monte-Carlo integrated colours of 500 realizations of 
        a 12~Gyr old GC populations with the labeled total masses and 
        [Fe/H] (see text for details). 
              }
         \label{fluct}
   \end{figure}
%

\subsection{Statistical fluctuations}

In Section~2 we already discussed the fact that when the
number of objects harboured by a stellar system is not
large enough to sample smoothly all evolutionary phases, statistical
fluctuations of its integrated colours do arise, caused by stochastic
variations of the number of objects populating the
faster evolutionary phases. 

The effect on the $(V-I)-(V-K)$ plane is shown by Fig.~\ref{fluct},
where we display the integrated colours obtained from 500 
realizations of a 12~Gyr old population with [Fe/H]=$-$1.62, and two
different values of the total mass $M_t$, namely $M_t =1.5 \ 10^5
M_{\odot}$ (that is the value corresponding to the peak of the GCLF) 
and $M_t =10^6 M_{\odot}$. For each
realization we have determined the integrated colours using the
Monte-Carlo procedure presented in Section~2. 

The points are centred around
the corresponding analytical values of the integrated colours, with a large
scatter. The total $(V-K)$ range is of the order of
$\approx$0.6~mag for $M_t =1.5 \ 10^5 M_{\odot}$, and $\approx$0.4~mag
for $M_t =10^6 M_{\odot}$. The total $(V-I)$ range is 
$\approx$0.2~mag for $M_t =1.5 \ 10^5 M_{\odot}$ and $\approx$0.1~mag
for $M_t =10^6 M_{\odot}$. There is a clear correlation between the
fluctuations of the two colours. 
The scatter decreases with increasing cluster mass, due to
a better sampling of the faster evolutionary phases when the cluster mass
(hence the total number of stars) increases. The $(V-K)$ colour
fluctuates more than $(V-I)$, because the integrated flux in the $K$-band
is more affected than the $I$-band by the AGB and bright RGB
evolutionary stages, that are the fastest ones and more prone to
statistical number fluctuations.
Due to these fluctuations only (we did not include any photometric
error in the simulations displayed in Fig.~\ref{fluct}) the [Fe/H]
value retrieved for an individual cluster can be wrong by as much as
$\approx$1~dex in the low metallicity regime. 
Retrieved ages can also be biased by several Gyr.

Figure~\ref{fluct} shows also the colour fluctuations associated to a 12~Gyr
GC with [Fe/H]=$-$0.6. It is important to notice that in this case the
distribution of points is almost parallel to the constant age line. This means
that at the upper end of the Galactic GC metallicity distribution, 
statistical colour fluctuations have a minor effect on the age
estimates using the $(V-K)-(V-I)$ diagram.

To study the effect of colour fluctuations on the retrieved age distribution 
of an entire GC system, instead of a single cluster 
-- including the presence of 0.03~mag 1$\sigma$
photometric errors -- we have determined one synthetic population 
corresponding to the 
template one, and one corresponding to the bimodal-age (12+4~Gyr) population, 
but this time the integrated colours of the individual
clusters are determined following the Monte-Carlo procedure explained
in Section~2 (these two populations are labelled 'population E' and 'F', respectively).        
The analysis of these two samples allows us to 
check, among others, whether statistical fluctuations can in principle erase the
difference between single-age and bimodal-age CADs. 
We stress that the inclusion of statistical fluctuations is
in principle necessary to realistically reproduce the integrated
photometry of observed GCs, unless they have a mass distribution
shifted towards masses much larger than the Galactic counterpart. 

\begin{figure}
   \centering
   \includegraphics[angle=0,width=9cm]{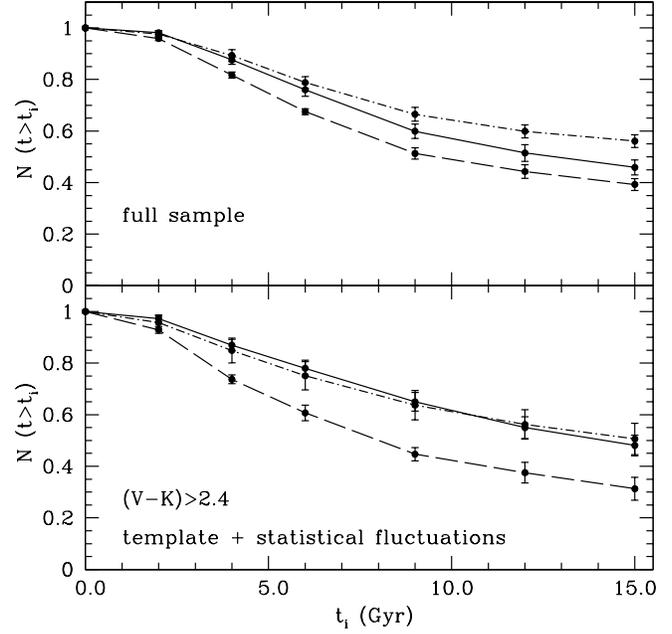}
      \caption{Comparison between the retrieved CADs for the 12~Gyr template
    (solid line) and population E, that accounts for statistical fluctuations 
   of the star number counts (dashed-dotted lines).
   The upper panel displays the comparison for the full synthetic samples, the
    lower panel the comparison only for objects with $(V-K)>$2.4. 
    The corresponding CADs for the bimodal-age population A with
   analytical integrated colours are also
    displayed as dashed lines.
              }
         \label{clf7}
   \end{figure}

\begin{figure}
   \centering
   \includegraphics[angle=0,width=9cm]{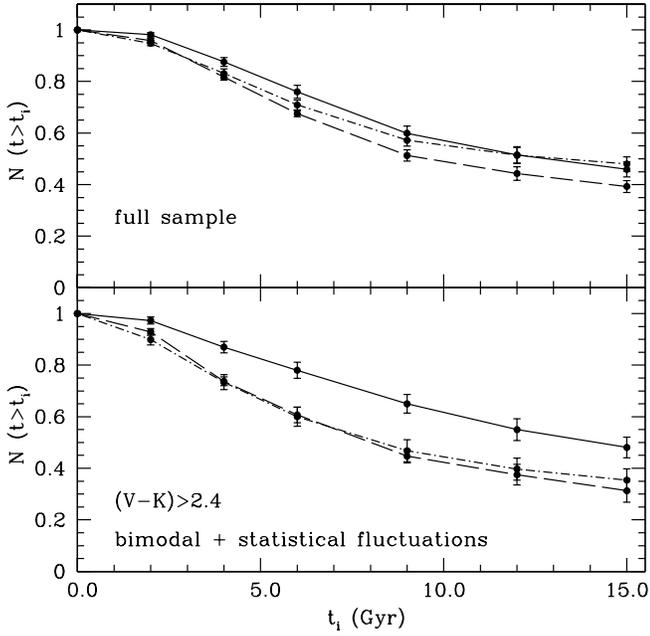}
      \caption{Comparison between the retrieved CADs for the 12~Gyr template
    (solid line) and the bimodal age population F,  
   with integrated colours computed via a Monte-Carlo simulation (dashed-dotted lines).
   The upper panel displays the comparison for the full synthetic samples, the
    lower panel the comparison only for objects with $(V-K)>$2.4. 
    The corresponding CADs for the bimodal-age population A with
   computed analytical colours are also
    displayed, as dashed lines.
              }
         \label{clf8}
   \end{figure}

In Figs.~\ref{clf7} and \ref{clf8} we have compared the CADs retrieved for
these two populations E and F, with those obtained for exactly same GC properties but 
with integrated colours computed analytically. 

%

The general trends are clear. When only objects with $(V-K)>$2.4 are
considered, the CADs are essentially unchanged compared to the case of 
analytical colours. This is a consequence of the fact that at high
metallicities the colour fluctuations do not significantly affect the
individual age estimates (see Fig.~\ref{fluct}). Very different is the case when
the whole samples are considered. The inclusion of the more metal
poor clusters causes a significant change of the CAD, compared to 
the case of analytical colours. Significant differences start between
the 6~Gyr and the 9~Gyr age bin. The CAD that accounts for the statistical fluctuations
is above the CAD for the analytical case, with a larger percentage of objects
at high ages. This behaviour is very similar for both the single-age (Fig.~\ref{clf7}) and 
bimodal-age (Fig.~\ref{clf8}) populations 

%

To summarize, the effect of statistical colour fluctuations on the 
retrieved CADs is negligible if only the redder clusters are analyzed. 
When the full GC samples are considered, a proper inclusion of statistical 
colour fluctuations tend to shift the retrieved CADs towards higher values 
-- for ages $t_i >$6~Gyr -- compared to the case when they are neglected. 
This behaviour is very similar for both the single-age (Fig.~\ref{clf7}) and 
bimodal-age (Fig.~\ref{clf8}) populations. As a consequence, 
relative differences between single-age and bimodal-age CADs are preserved.

\begin{figure}
   \centering
   \includegraphics[angle=0,width=9cm]{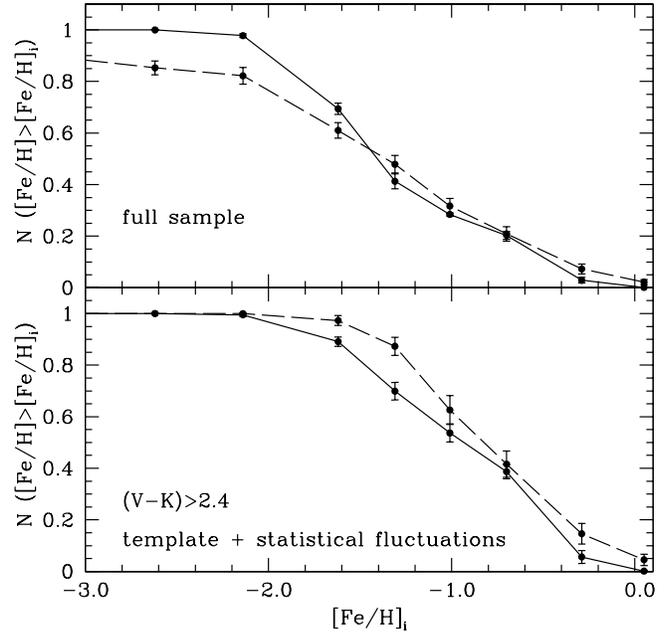}
      \caption{As in Fig.~\ref{mlf1} but for the 12~Gyr old population E 
   that accounts for colour statistical fluctuations. 
              }
         \label{mlf6}
   \end{figure}

\begin{figure}
   \centering
   \includegraphics[angle=0,width=9cm]{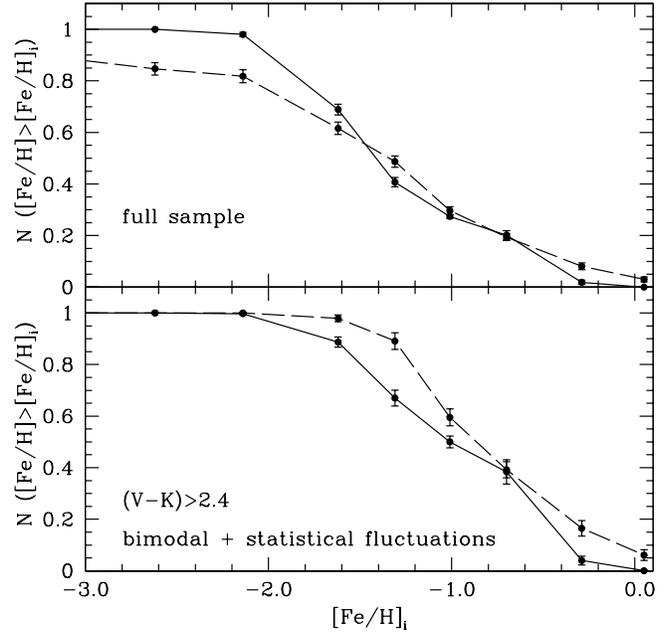}
      \caption{As in Fig.~\ref{mlf1} but for the bimodal-age population F 
   that accounts for colour statistical fluctuations. 
              }
         \label{mlf7}
   \end{figure}

As for the [Fe/H] distribution,  
the retrieved CMeDs that include the effect of colour 
fluctuations are displayed 
in Figs.~\ref{mlf6} and \ref{mlf7}, compared to the intrinsic CMeDs. 
First, one can immediately notice that, again, the assumed age distribution 
does not play a significant role in the outcome of this comparison.  
In general, the number fractions for the full samples (with the exception of the
point at [Fe/H]=$-$4.5, that by definition is always equal to unity for both
intrinsic and retrieved CMeDs) are typically lower than the 
intrinsic CMeDs when [Fe/H] is below $\sim -$1.6. This is due to a
sizable fraction of metal poor objects with anomalously blue $(V-K)$ 
colours, that are assigned a [Fe/H]$< -$2.62.  
At higher metallicities the differences with respect to the intrinsic
CMeDs tend to be smaller, and only slightly
larger than in case of photometric errors only 
(see Figs.~\ref{mlf1} and \ref{mlf2}).

%

As an interesting detail, one can notice that the intrinsic [Fe/H] 
(solid lines in Figs.~\ref{mlf6} and \ref{mlf7}) 
of objects with $(V-K)>$2.4 can reach
values even lower than [Fe/H]=$-$1.62. In fact, the intrinsic CMeD corresponding 
to ${\rm [Fe/H]}_i> -$1.62  displays a value below
unity; if all clusters in the sample have an intrinsic [Fe/H] higher than this value, the corresponding 
CMeD would show a value equal to unity at this point.
Why do we find such metal poor clusters in the red subsample?   
This occurrence is not mainly due to the photometric error employed in the
simulation -- Figures~\ref{mlf1} and \ref{mlf2} display values
approximately equal to unity at this [Fe/H] -- rather to 
the colour fluctuations of the most metal poor
objects, that can appear at anomalously blue but also anomalously high $(V-K)$ colours 
(see Fig.~\ref{fluct}). As a consequence the intrinsic CMeD  
of the $(V-K)>$2.4 subsamples will contain anomalously metal poor objects. Another consequence 
is that the retrieved metallicity of these anomalously red objects is 
much higher than their intrinsic one, when estimated from their position in the colour-colour plane.  
This explains why the retrieved CMeD at 
${\rm [Fe/H]}_i= -$1.62 displays a value much closer to unity, compared to the intrinsic one.  

%

Before closing this discussion on colour fluctuations, we wish to
address a further important point. Observations of distant GC systems usually
can acquire sufficiently precise photometry of only the brightest
clusters, populating the bright tail of the GCLF. Brightest
clusters means the most massive ones, for which the effect of
statistical colour fluctuations is smaller. To simulate this effect
and check the consequences on CAD and CMeD, we selected from our
populations E and F  only objects with integrated $M_V < -$8.5 
that corresponds to $M_t$ larger than
$\sim 3 \ 10^5~M_{\odot}$. The resulting samples contain about 60
objects, therefore the error bars on the individual points of the
retrieved CADs and CMeDs will be larger than in the previous case. 
 
   \begin{figure}
   \centering
   \includegraphics[angle=0,width=9cm]{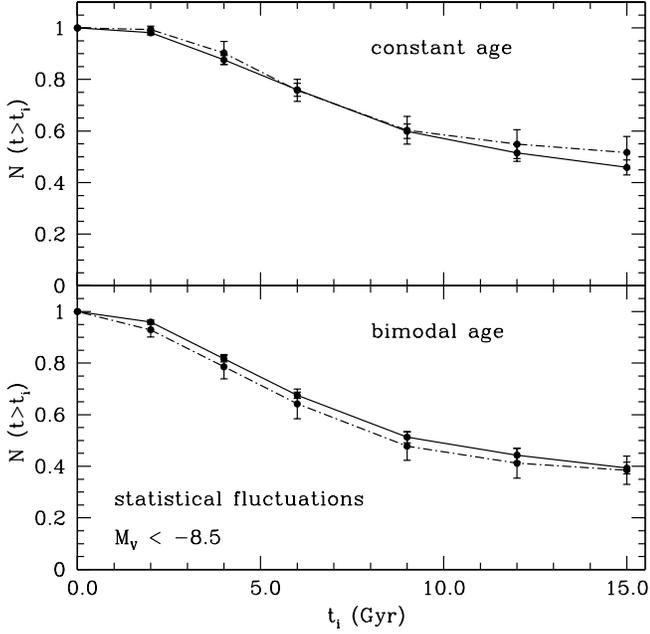}
      \caption{Retrieved CADs for the 12~Gyr population E
    (upper panel) and the bimodal-age population F (lower panel),   
   with integrated colours computed via a Monte-Carlo technique. Dashed-dotted
   lines represent the case of GC subsamples with integrated $M_V< -$8.5
   (about 60 clusters for each Monte-Carlo realization). 
   Solid lines display the CADs retrieved for the full samples 
   corresponding to the template and population A,  
   with integrated colours computed analytically (see text for details).
               }
         \label{clf9}
   \end{figure}
%

Figure~\ref{clf9} compares the retrieved CADs of the single-age population E 
and bimodal-age population F (that include colour fluctuations) restricted to
$M_V< -$8.5, with the results for the full template and population A. The
effect of statistical fluctuations on the ages is now almost
negligible. 

\begin{figure}
   \centering
   \includegraphics[angle=0,width=7cm]{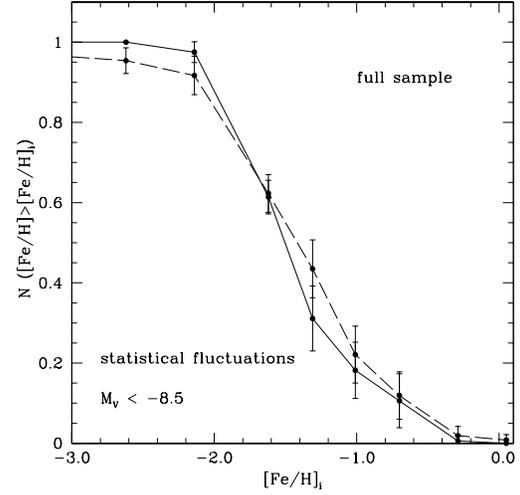}
      \caption{As in the upper panel of Fig.~\ref{mlf7} but for the
   sample restricted to objects with  $M_V< -$8.5 (see text for details).
              }
         \label{mlf8}
   \end{figure}

The retrieved CMeD of the population E sample ( restricted to
$M_V< -$8.5) is compared to the 
intrinsic one in Fig.~\ref{mlf8} (analogous result is obtained for
the bimodal-age sample) and shows at low [Fe/H] much 
smaller differences compared to the intrinsic  CMeD than the case
displayed in Fig.~\ref{mlf6}, because of the reduced 
fluctuations\footnote{The two distributions appear 
formally in agreement within the 1$\sigma$ error bars, but notice the 
larger errors -- due to the relatively small number of
objects -- associated to the $M_V < -$8.5 sample.}. 

As a general conclusion of this further test, colour fluctuations do 
not play a relevant role in the determination of the age and [Fe/H] 
distributions from the integrated $(V-I)-(V-K)$ diagram, when one 
restricts the analysis to GCs brighter than $M_V = -$8.5.

%

\subsection{Core overshooting}

There is an ongoing debate about 
the possibility that in stars with convective cores 
along the Main Sequence (i.e., with mass 
larger than $\sim$ 1.1- 1.2$M_{\odot}$) the motion 
of the convective fluid elements is not halted shortly beyond the boundary with
the stable radiative region as determined by the Schwarzschild
criterion, but it goes on for an appreciable length within the
formally stable surrounding layers (overshooting). 
The extent of the overshooting region is the focus of many
investigations like, among the most recent ones, those by Testa et
al.~(\cite{tfc99}), Woo \& Demarque~(\cite{wd01}) and  
Barmina, Girardi \& Chiosi~(\cite{bgc02}).
The presence of efficient overshooting from Main Sequence convective
cores alters both colour and luminosity of the isochrones Turn Off (such that 
the Turn Off is generally brighter and hotter at fixed age) thus
affecting integrated colours of an SSP of fixed age and
chemical composition.

This issue is irrelevant for Galactic globular clusters, given that 
they harbour stars with mass low enough not to experience the onset 
of convective cores along the Main Sequence phase; however, in 
presence of bimodal-age populations with a component only a few Gyr
old, this phenomenon can play a role. 

   \begin{figure}
   \centering
   \includegraphics[angle=0,width=9cm]{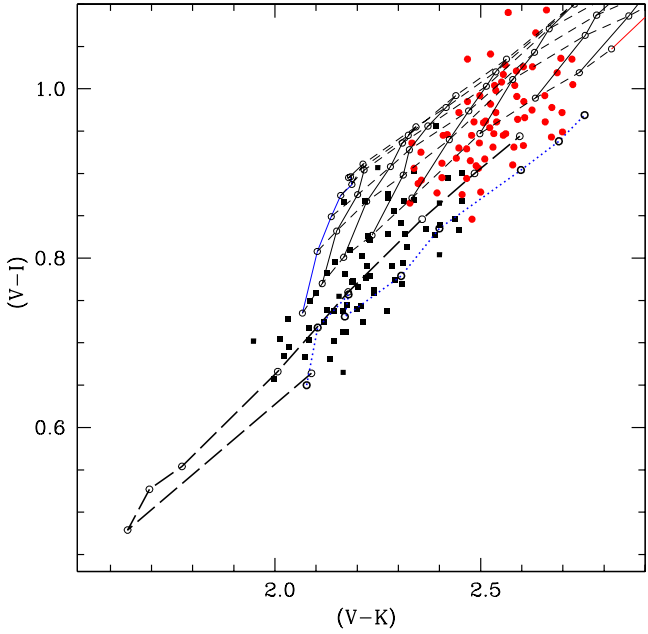}
      \caption{Colour-colour diagram of two 2~Gyr old synthetic GC
   populations (80 clusters each) 
   with [Fe/H]=$-0.55 \pm 0.20$, computed using models with (black filled
   squares) and without (red filled circles) core overshooting. The reference
   colour grid of Fig.~\ref{template} is also displayed, together with
   the 2~Gyr constant age line obtained from models that include
   overshooting (thick dashed line) and the 1~Gyr constant age sequence from models
   without overshooting (dotted line -- see text for
   details). The [Fe/H] values decrease moving away from the 
   point with the largest $(V-K)$ value, along the constant age line.
               }
         \label{grid2}
   \end{figure}
%

Figure~\ref{grid2} displays our reference $\alpha$-enhanced
colour grid shown in 
Fig.~\ref{template}, obtained from models without overshooting, with
the addition of the 1~Gyr old sequence. 
Colours predicted by
$\alpha$-enhanced models including overshooting (calculated
from isochrones in the \textit{BaSTI}
database, that employ the overshooting  treatment discussed by  
Pietrinferni et al.~\cite{pietr04}) are the same as the 
reference grid, when the
age is equal or larger than 4~Gyr. At younger ages convective
overshooting starts to play a role. For the sake of comparison we
display in Fig.~\ref{template} also the 2~Gyr constant age line
obtained from models including overshooting. 

Quantitative and qualitative differences between the 2~Gyr lines with
and without overshooting
are remarkable. On the whole the 2~Gyr overshooting colour-colour sequence lies
between the 1~Gyr and 2~Gyr non overshooting ones.  
More specifically, for a given age and [Fe/H] pair, 
the $(V-I)$ and $(V-K)$ colours of the 2~Gyr overshooting models are
systematically bluer than the values of the reference
grid at 2~Gyr; the trend of both colours with [Fe/H] at constant age is
non-monotonic in the overshooting case, because of the turn to the red
(in both $(V-K)$ and $(V-I)$) occurring between the two lowest metallicity points. The 1~Gyr
non-overshooting line shows also signs of a non-monotonic behaviour at low
metallicities. These are due to the complex interplay between the 
variation of Turn Off colours, colour extension of the blue loops during the central He-burning
phase and brightness of the RGB phase with changing [Fe/H].  

   \begin{figure}
   \centering
   \includegraphics[angle=0,width=9cm]{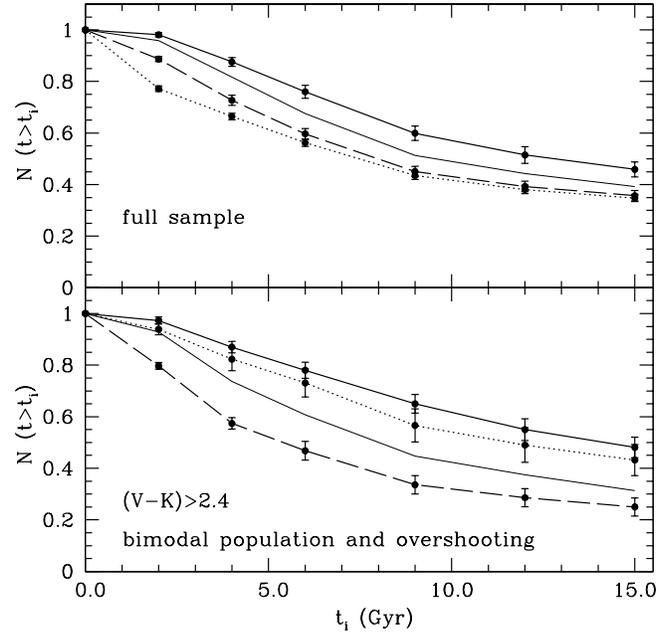}
      \caption{CADs retrieved using the reference
   non-overshooting color-colour calibration of Fig.~\ref{template}
   for the 12+2~Gyr bimodal-age populations H (dotted line) and G (dashed
   line). The CAD retrieved for the template (thick solid line) and the bimodal-age 
   population A (thin solid line without symbols) are
   also displayed. The upper panel displays the CADs for the full
   samples, the lower panel the case of objects with $(V-K)>$2.4.
               }
         \label{clf10}
   \end{figure}
%

To highlight further the effect of convective overshooting on
integrated colours, 
Fig.~\ref{grid2} shows also the colour-colour diagram of two synthetic
samples (80 objects each) of 2~Gyr old clusters with [Fe/H]=$-0.55 \pm 0.20$ 
(i.e. the [Fe/H] distribution of the metal rich components of all
synthetic populations discussed in this paper) and 1$\sigma$ 
photometric errors of 0.03~mag, 
computed from $\alpha$-enhanced $\eta$=0.2 models both with and without 
convective overshooting during
the Main Sequence phase. The inclusion of overshooting, as expected,
makes the synthetic clusters bluer on average in both colours. This 
certainly affects the retrieved CADs and CMeDs, if an inconsistent 
colour-colour calibration is used. 

For a quantitative assessment of this effect 
we have created a bimodal-age population with the same [Fe/H] distribution as 
the template sample, a 12~Gyr age for the metal poor component, and a
2~Gyr age for the metal rich one. Photometric errors are as in the
template sample. We computed the integrated colours analytically using
the $\eta$=0.2 $\alpha$-enhanced isochrones, both with (this population is 
labelled 'population H')  
and without (this population is labelled 'population G') overshooting
for the case of the 2~Gyr clusters. 

Figure~\ref{clf10} displays the CADs retrieved using the reference
non-overshooting color grid for both populations G and H,
compared to the results for the template sample, and for the bimodal 
(12+4~Gyr) population A. 
   \begin{figure}
   \centering
   \includegraphics[angle=0,width=9cm]{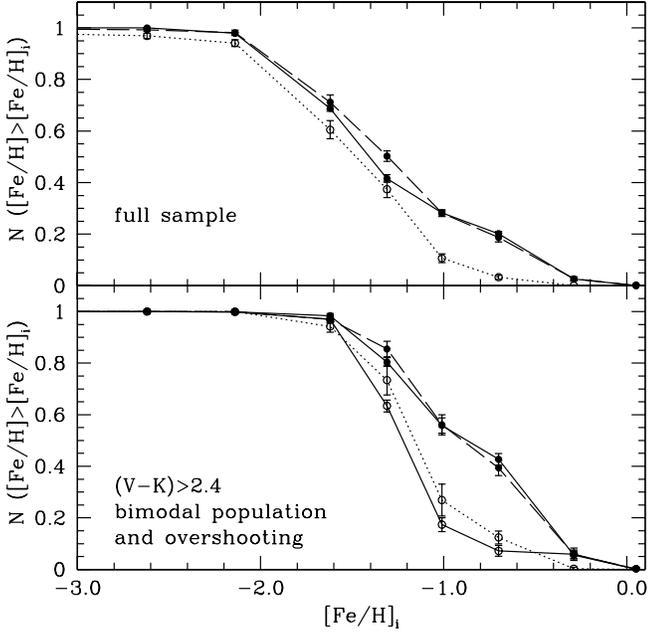}
      \caption{CMeDs retrieved for populations G and H. 
   Dashed lines and filled symbols
   correspond to population G, dotted lines and open
   symbols to population H. Solid lines with filled and open
   symbols are the input CMeDs for populations G and H, respectively.
   The top panel shows the comparison for the full samples, the bottom
   panel the case of objects with $(V-K)>$2.4.
               }
         \label{mlf9}
   \end{figure}
%

We first discuss the comparison between the two 12+2~Gyr
populations H and G, with and without overshooting.  
When the full samples are considered, the sample that includes 
overshooting displays a CAD differing from the non-overshooting
one only in the youngest bins. This is easy to understand from 
Fig.~\ref{grid2}; the ``overshooting'' 2~Gyr old population is 
located almost always below the 2~Gyr line of the
reference colour grid, whereas the ``non-overshooting''
population of the same age is -- obviously -- more evenly distributed
around the 2~Gyr line of the reference grid.  
This explains the appreciable difference between the two CADs in the bins
corresponding to objects with ages larger than 2 and
4~Gyr, respectively. Differences decrease with increasing age because the
contribution of the young population decreases sharply, and what
dominates is the 12~Gyr old component which is unaffected by
overshooting. 

When we restrict the comparison to objects redder than $(V-K)$=2.4, the
differences are more extreme. In this colour range there
are very few 2~Gyr old clusters 
for population H (see again Fig.~\ref{grid2}) 
so that the large majority of objects belong to the old
metal poor 12~Gyr component. The CAD of population G  
shows predictably much lower number fractions, because of the
presence of 2~Gyr old objects in addition to the 12~Gyr old
component. 
Also, the number of clusters in this colour
range is only about 25\% of the total sample for the overshooting
case, whereas the same fraction 
is about 50\% for the non-overshooting sample. This smaller
sample size explains the larger error bars attached to population H 
CAD when compared with population G.

We now compare the CAD of these two 12+2~Gyr populations with our 12~Gyr template 
sample and the bimodal population A with 12+4~Gyr old subpopulations. 
As expected, the CADs of 
both 12+2~Gyr systems have systematically lower values
at each age bin than the CADs of both the template and population A. 
In case of objects redder than $(V-K)$=2.4 the CAD of population H 
population is now very close to the template one, because it contains
essentially only 12~Gyr old objects. This means that an analysis
restricted to this colour range would assign to this bimodal 
population an essentially constant and old age.  
On the other hand the CAD of the non overshooting population G   
shows, as expected, lower number fractions than the template and population A, even 
when the analysis is restricted to the redder clusters. 

We can now summarize the effect of Main Sequence convective core 
overshooting on the retrieved CAD. When a colour grid that does not
include core overshooting is used to retrieve ages of a young 
component where overshooting is efficient, the retrieved CAD 
is biased towards young ages in the age interval below $\sim$ 4~Gyr. 
If the analysis is restricted to the redder clusters, the retrieved CAD 
lacks almost completely the young component. An interesting consequence is 
that when one restricts the analysis to objects redder than a given
$(V-K)$, there is the possibility to overlook the presence of a young component 
if the treatment of core convection in the models is not adequate.  

Turning now to the estimate of the [Fe/H] distributions, 
the CMeDs retrieved for the two bimodal-age populations G and H are displayed in
Fig.~\ref{mlf9}. The CMeD of population G shows only
minor deviations from the input one, whereas in case of population H 
the most evident difference is the drop 
between ${\rm [Fe/H]}_i =-$1.31 and $-$1.01 when the full sample is
considered. This is due to the 
systematically lower [Fe/H] 
estimated for the metal rich 2~Gyr component (see Fig.~\ref{grid2}) 
whose retrieved [Fe/H] has in fact a sharp drop around [Fe/H]$\sim -$1. 
When the analysis is restricted to objects redder than $(V-K)$=2.4, population H  
is almost completely lacking the young metal rich 
component, located at bluer colours. Compared to the case of population G, 
both input and retrieved CMeDs display a large drop between ${\rm [Fe/H]}_i =-$1.62
and $-$1.01, because they are sampling essentially the
high-metallicity tail of the metal poor old subpopulation.

\subsection{Varying the 1$\sigma$ photometric error}

For a given $M_t$, [Fe/H] and age distribution of the synthetic GC
systems, the magnitude of the photometric error can affect the
quantitative results of our comparisons (see the discussion in 
Hempel et al.~\cite{hem04}). 
We briefly investigate here what happens if we 
consider photometric errors larger than our assumed value of
0.03~mag. We examine the cases of the template 12~Gyr 
sample, and the 12~Gyr $\eta=0.4$ opopulation B, discussed in
Section~\ref{eta}. This latter sample is the one whose 
retrieved CAD appears to be the most discrepant 
from the template one (see upper panel of Fig.~\ref{clf4}), approximately 
comparable to a bimodal 12+4~Gyr population A with $\eta$=0.2. 

\begin{figure}
   \centering
   \includegraphics[angle=0,width=9cm]{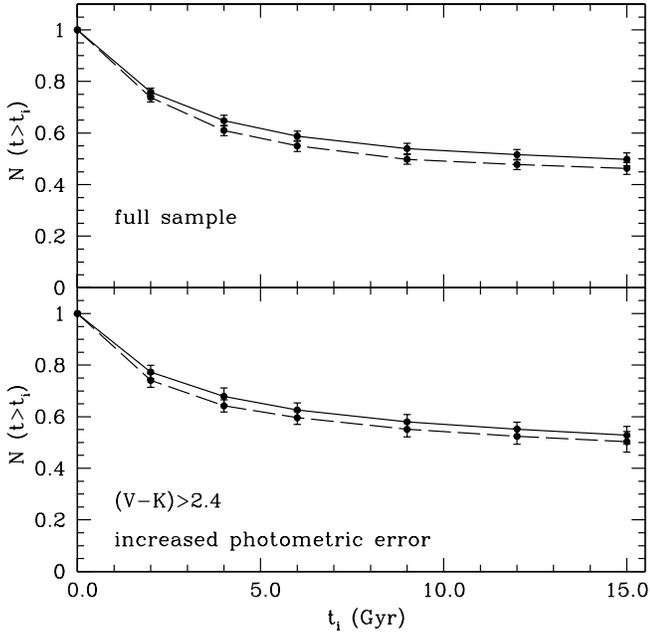}
      \caption{Comparison between the retrieved CADs for population I and J 
    (see text and Table~\ref{table1} for details) that include  
    1$\sigma$ photometric errors equal to 
    0.10~mag . The upper panel displays
   the comparison considering the full cluster samples, the lower panel the
   comparison restricted to clusters with $(V-K)>$2.4.
               }
         \label{clf11}
   \end{figure}

Figure~\ref{clf11} displays the CAD retrieved from a population 
(labelled 'population I') 
analogous to the 12~Gyr template, but including a 1$\sigma$ photometric error equal to
0.10~mag. Compared to the template CAD of Fig.~\ref{clf1} one can notice a
steeper slope at young ages and a flatter one at older ages. This
reflects the much larger dispersion of points in the colour-colour diagram,
with sizable numbers of clusters younger than 2~Gyr or older than 15~Gyr.  
%
We have plotted in the same figure also the CADs retrieved by applying our
reference grid to a population 
(labelled 'population J') analogous to the 12~Gyr GC sample with $\eta$=0.4, 
but including  
again a 1$\sigma$ photometric error equal to 0.10~mag.  
Comparisons with population I show that the retrieved CADs are
almost identical when considering both the whole populations or the 
redder clusters only. 

This result shows that 1$\sigma=$0.10~mag photometric errors are sufficient 
to  erase all the systematic effects (the conclusion for the CMeDs is similar) 
discussed in this paper. Given the similitude between the 
CADs retrieved for a 12~Gyr old GC sample with $\eta$=0.4 
and the bimodal 12+4~Gyr population with $\eta$=0.2 (see Section~\ref{eta}) 
Fig.~\ref{clf11} implies also 
that bimodal age populations with age
differences of up to 8~Gyr go undetected from CAD analyses, in presence 
of photometric errors of this magnitude. 

\section{Summary and discussion}

In the previous section we have extensively studied the effect of 
inconsistent HB morphology, metal mixture and treatment of core
convection between a synthetic sample of GCs (corresponding to an
hypothetical observed system) and the reference colour grid used to
retrieve their age and [Fe/H]. We also investigated the role played by
statistical colour fluctuations due to the low $M_t$ of GCs. 
For all our tests we have created synthetic GC samples with $M_t$ 
(composite power law with two different slopes) and
[Fe/H] (bimodal metal rich + metal poor components) 
distributions appropriate for the Galactic GC system, and a
1$\sigma$ photometric error typical of the integrated colours of
Galactic globulars. The photometric error only is able to 
transform a constant or bimodal age distribution into an age
distribution that covers smoothly a larger age range, as shown 
already by Hempel \& Kissler-Patig~(\cite{hem04}) . 
In addition to this smoothing, due purely to photometric errors, the
results of our analysis can be summarized as follows:

\begin{itemize}
    \item All effects studied in this paper may modify, to various
    degrees, the GC age distributions
    obtained from integrated $(V-K)-(V-I)$ diagrams. The retrieved
    [Fe/H] distributions are overall less affected.
    \item An HB morphology bluer than the one adopted for the
    reference colour grid used to estimate the GC ages 
    can simulate a bimodal-age population 
    (old + young components) even if all GCs are coeval. 
    This is true both in case of a bluer morphology 
    for the whole cluster sample, or when a substantial fraction of 
    clusters has a bluer HB than accounted for in the theoretical colour grid. 
    For the mass loss laws employed in our analysis this effect is negligible when
    only the red clusters are considered (in our case clusters with
    $(V-K)>$2.4). More extreme amounts of mass lost along the RGB may
    influence also the ages of red clusters.  
    Biases in the retrieved [Fe/H] distributions compared to the
    intrinsic ones are small, and restricted mainly to low [Fe/H],
    with an anomalously high number of objects assigned [Fe/H] values below
    $\sim -$2.0.
    \item In the presence of a scaled-solar metal rich GC component
    (the effect of changing the metal mixture is larger with
    increasing [Fe/H]) the use of an $\alpha$-enhanced metal mixture in the reference colour
    grid introduces an age bias, but the retrieved [Fe/H] distribution
    is unaffected. In our tests 
    the retrieved age distribution tends to simulate a bimodal-age
    sample -- to a lesser degree than the blue HB tests -- 
    even if all GCs have the same age. The effect is more pronounced  
    when subsamples containing only the redder (more metal rich) clusters are considered.    
    \item The effect of statistical colour fluctuations, if
    unaccounted for in the age estimate, does not affect the age
    distribution of red clusters; it tends overall to increase the
    number of objects with retrieved ages larger than 
    6 -- 9~Gyr, when also the metal poor clusters are considered. 
    The estimated [Fe/H] distributions are affected to different
    degrees, whether the whole sample or only red clusters are considered. 
    In the former case the metal poor clusters tend to be assigned an  
    underestimated [Fe/H], whereas in the latter case there is the tendency
    to overestimate [Fe/H].   
    When $M_t$ is above $\approx 3 \ 10^5 M_{\odot}$ (integrated $M_V
    > -$8.5) the effect of colour fluctuations on the retrieved age (and [Fe/H]) 
    distribution of the synthetic GC samples is almost negligible.
    \item The effect of Main Sequence core overshooting is relevant
    only for ages below $\sim$4~Gyr. If a colour grid that does not
    include core overshooting is used to retrieve ages of an old+young composite 
    population where overshooting is efficient, the resulting age
    distribution of the young component will be biased towards low ages. Also the 
    retrieved [Fe/H] values will be lower than the true ones (the opposite is of course
    true if the reference grid includes overshooting but this is not
    efficient in the stars belonging to the GC population).  
    Subsamples containing only the redder clusters could lack 
    almost completely the young component. 
    \item If the 1$\sigma$ photometric error on the cluster
    photometry reaches $\sim$0.10~mag, all systematic effects  
    found in our analysis are erased. Photometric errors of this size also prevent the
    detection of bimodal (old + young) age populations with age differences of 
    $\sim$8~Gyr.
\end{itemize}

We conclude this section with a few additional considerations on GC age distributions 
obtained from integrated colour-colour diagrams. One of our conclusions is that 
1$\sigma$ photometric errors of the order of 0.1~mag would prevent the identification of bimodal 
populations with age differences of $\sim$8~Gyr. A way to mitigate this problem is to make use 
of colours with a wider dynamical range. The $(U-I)-(V-K)$ diagram would greatly 
improve in this respect (see Hempel \& Kissler-Patig~\cite{hem04b}). 
To this purpose, in the Appendix we provide accurate fitting formulae 
$(U-I)_{t_i}=f[(V-K)]$ (for the $\eta$=0.2 grid) that allow a fast estimate of
the age distribution from this colour-colour diagram. 

However, the use of the $(U-I)$ colour instead of $(V-I)$ does 
not avoid the other systematics discussed in this paper, in particular 
the thorny problem of the unknown HB morphology in unresolved systems.

   \begin{figure}
   \centering
   \includegraphics[angle=0,width=8cm]{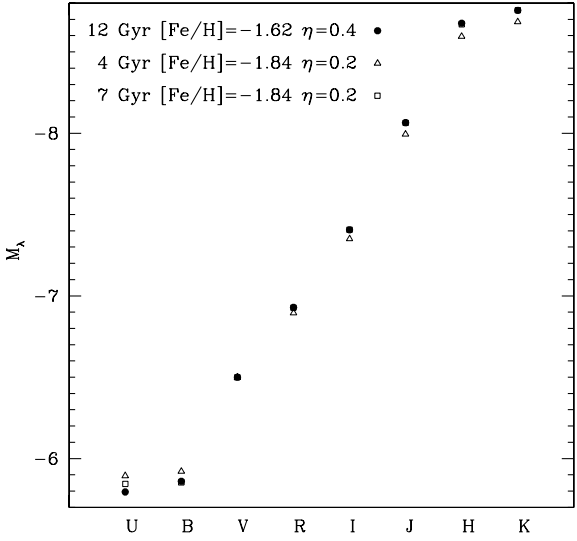}
      \caption{Comparison of the SED ($UBVRIJHK$ bands) of three synthetic GCs, determined from 
$\alpha$-enhanced isochrones with the labelled age, metallicity  
and $\eta$ values. The integrated magnitudes are normalized in such a way that 
$M_V=-6.5$ (an arbitrary value) for all three cases. No random photometric errors are included. 
               }
         \label{SEDcomp}
   \end{figure}


Making use of both $(U-I)-(V-K)$ and $(V-I)-(V-K)$ diagrams 
might, under certain conditions, help to detect the presence of an HB population 
with a different morphology compared to the 
calibrating colour grid. The reason is explained by Fig.~\ref{SEDcomp}. Here 
we show the SED (integrated broadband magnitudes from $U$ to $K$) for three different 
synthetic GCs computed with varying ages, [Fe/H] and mass loss parameter $\eta$.  
After normalization to the same (arbitrary) value of $M_V$, one can notice the following. 
First of of all, all magnitudes 
between $B$ and $K$ are identical for a 12~Gyr [Fe/H]=$-$1.62 population with $\eta$=0.4, and a 7~Gyr, 
[Fe/H]=$-$1.84 population with $\eta$=0.2. As a consequence, all colour combinations using filters 
from $B$ to $K$ (i.e. the $(V-I)-(V-K)$ diagram) show the same degeneracy, and the 12~Gyr $\eta$=0.4 
population is assigned a 7~Gyr age from a $(V-I)-(V-K)$ analysis that employs 
the reference $\eta$=0.2 grid. 
Moreover, when 0.03~mag 1$\sigma$ random photometric errors employed in our tests are applied to 
the magnitudes of the 12~Gyr system, the $(V-I)$ and $(V-K)$ colours of the 4~Gyr, 
$\eta$=0.2 population (and even lower ages) are well within reach. This kind of comparison 
explains the results displayed in Fig.~\ref{clf4}.
 
The normalized $U$ magnitude is however different between the 12 and the 7~Gyr populations, 
implying that colours involving the $U$-band 
(i.e. $(U-I)$) are not coincident. The consequence for the age estimates is that 
if one employs both $(U-I)-(V-K)$ and $(V-I)-(V-K)$ diagrams, an inconsistent HB morphology 
between observed population and theoretical grid, 
could potentially manifest itself through different age distributions obtained 
from the two colour-colour diagrams. 
Unfortunately, the difference between the normalized $U$ magnitudes 
of the 7~Gyr and the 12~Gyr populations is equal to only 0.05~mag (it is of 0.10~mag 
between the displayed 4~Gyr and 12~Gyr populations). 
The presence of 1$\sigma$ photometric errors of, e.g. 0.03~mag, 
tends to smear out this difference and would make the use of the $U$-band 
probably less helpful. 
These figures are clearly dependent on the exact morphology of the HB 
in the unresolved populations. Larger differences -- compared to our test -- between the 
HB morphologies of the observed population and the theoretical grid may 
still make the $(U-I)-(V-K)$ and $(V-I)-(V-K)$ CAD analysis a very useful diagnostic of 
the presence of blue HB objects unaccounted for in the theoretical analysis.

\begin{acknowledgements}
We warmly thank Phil A. James for a preliminary reading of the manuscript, 
insightful comments and suggestions, and an anonymous referee for constructive comments that 
helped us to improve the paper. S.C. acknowledges partial financial
support from INAF and MIUR.

\end{acknowledgements}

\appendix
\section{Useful analytical fits for determining Cumulative Age Distributions}

The CADs presented in this paper have been computed by 
associating to each individual cluster an age greater than $t_i$, when
it lies above the line of constant $t_i$ in the $(V-I)-(V-K)$ plane. 
Lines of constant $t_i$ for our reference 
$\alpha$-enhanced, non-overshooting isochrones with $\eta$=0.2, are well
represented -- within 0.01~mag in $(V-I)$ and 0.025~mag in $(U-I)$ - 
by the following analytical relationships (the range of validity of the fits is also given): 
$$(V-I)_{2~Gyr} = 0.9807 \ {\rm ln}[(V-K)] + 0.036 $$
$$(U-I)_{2~Gyr} = 3.1552 \ {\rm ln}[(V-K)] - 1.097 $$
$$2.06 < (V-K) < 2.82 $$
$$(V-I)_{4~Gyr} = 0.8963 \ {\rm ln}[(V-K)] + 0.148 $$
$$(U-I)_{4~Gyr} = 3.1487 \ {\rm ln}[(V-K)] - 1.022 $$
$$2.10 < (V-K)  < 2.96$$
$$(V-I)_{6~Gyr} = 0.8605 \ {\rm ln}[(V-K)] + 0.198 $$
$$(U-I)_{6~Gyr} = 3.2728 \ {\rm ln}[(V-K)] - 1.106 $$
$$2.13 < (V-K)  < 3.02$$
$$(V-I)_{9~Gyr} = 0.8390 \ {\rm ln}[(V-K)] + 0.233 $$
$$(U-I)_{9~Gyr} = 3.3364 \ {\rm ln}[(V-K)] - 1.132 $$
$$2.16 < (V-K) < 3.10$$
$$(V-I)_{12~Gyr} = 0.8572 \ {\rm ln}[(V-K)] + 0.224 $$
$$(U-I)_{12~Gyr} = 3.4256 \ {\rm ln}[(V-K)] - 1.197 $$
$$2.18 < (V-K)  < 3.18$$
$$(V-I)_{15~Gyr} = 0.8781 \ {\rm ln}[(V-K)] + 0.210 $$
$$(U-I)_{15~Gyr} = 3.4613 \ {\rm ln}[(V-K)] - 1.208 $$
$$2.18 < (V-K) < 3.25$$

When overshooting is included, the 2~Gyr sequences displays a non
monotonic behaviour for [Fe/H]$< -$2.14, in both $(V-I)$ and $(U-I)$. 
When [Fe/H]$\geq -$2.14,
the sequences are  well reproduced (within 0.015~mag in $(V-I)$ and 
0.03~mag in $(U-I)$) by: 

$$(V-I)_{2~Gyr, ov} = 1.0048 \ {\rm ln}[(V-K)] - 0.018 $$
$$(U-I)_{2~Gyr, ov} = 2.5858 \ {\rm ln}[(V-K)] - 0.611 $$
$$1.64 < (V-K) < 2.60 $$
corresponding to [Fe/H] between +0.05 and $-$2.14. 
The colours at [Fe/H]=$-$2.62 are $(V-K)$=2.09, $(V-I)$=0.66 and $(U-I)$=0.98.

Lines of constant $t_i$ for our  
scaled-solar non-overshooting isochrones with $\eta$=0.2 are well
represented (within 0.015~mag in $(V-I)$) by: 
$$(V-I)_{2~Gyr} = 0.8356 \ {\rm ln}[(V-K)] + 0.146 $$
$$(U-I)_{2~Gyr} = 3.0462 \ {\rm ln}[(V-K)] - 0.999 $$
$$2.05 < (V-K) < 2.91 $$
$$(V-I)_{4~Gyr} = 0.7958 \ {\rm ln}[(V-K)] + 0.226 $$
$$(U-I)_{4~Gyr} = 1.2806 \ (V-K) - 1.341 $$
$$2.11 < (V-K)  < 3.05$$
$$(V-I)_{6~Gyr} = 0.7577 \ {\rm ln}[(V-K)] + 0.283 $$
$$(U-I)_{6~Gyr} = 1.2893 \ (V-K) - 1.332 $$
$$2.15 < (V-K)  < 3.17$$
$$(V-I)_{9~Gyr} = 0.7617 \ {\rm ln}[(V-K)] + 0.297 $$
$$(U-I)_{9~Gyr} = 1.3492 \ (V-K) - 1.455 $$
$$2.17 < (V-K) < 3.26$$
$$(V-I)_{12~Gyr} = 0.7734 \ {\rm ln}[(V-K)] + 0.296 $$
$$(U-I)_{12~Gyr} = 1.3500 \ (V-K) - 1.425 $$
$$2.18 < (V-K)  < 3.35$$
$$(V-I)_{15~Gyr} = 0.8056 \ {\rm ln}[(V-K)] + 0.269 $$
$$(U-I)_{15~Gyr} = 1.3617 \ (V-K) - 1.438 $$
$$2.17 < (V-K) < 3.43$$

The 2~Gyr constant age sequences that include overshooting 
display the same monotonic behaviour as for
the older ages, and are well approximated (within 0.02~mag in
$(V-I)$ and 0.035~mag in $(U-I)$) by: 
$$(V-I)_{2~Gyr, ov} = 0.9194 \ {\rm ln}[(V-K)] + 0.010 $$
$$(U-I)_{2~Gyr, ov} = 1.2592 \ (V-K) - 1.402 $$
$$1.59 < (V-K) < 2.62 $$

We conclude this section with some further remarks about our reference 
$(V-I)-(V-K)$ calibration detailed above. 
We have compared the [Fe/H]-$(V-I)$ 
and [Fe/H]-$(V-K)$ empirical relationships obtained by 
Barmby et al.~(\cite{bar00}) from a sample of Galactic GCs, 
with the theoretical counterpart derived from our template 
12~Gyr sample, but with colours computed via a Monte-Carlo procedure, 
as appropriate when the spectrum of GC masses is considered. 
We recall that in this synthetic sample we have 
used a [Fe/H] distribution typical of Galactic GCs, and 
the photometric errors are consistent
with the errors on the colours employed by Barmby et
al.~(\cite{bar00}). Also the GC mass
distribution is consistent with the Galactic GCLF. An age of 12~Gyr
is typical of Galactic GC ages, as estimated by 
Salaris \& Weiss~(\cite{sw02}).

Barmby et al.~(\cite{bar00}) derive from empirical [Fe/H] values 
and integrated colours the following relationships:
$${\rm [Fe/H]}=(4.22\pm0.39) \ (V-I) - (5.39\pm0.35) \ \ {\rm 75 \ objects}$$
$${\rm [Fe/H]}=(1.40\pm0.17) \ (V-K) - (4.62\pm0.36) \ \ {\rm 35 \ objects}$$
We considered 30 realizations of our synthetic GC sample, 
and determined from each sample slopes and zero points of the [Fe/H]-$(V-I)$ 
and [Fe/H]-$(V-K)$ relationships. The mean values and 1$\sigma$
spreads around the mean provide:
$${\rm [Fe/H]}=(3.87\pm0.07) \ (V-I) - (5.14\pm0.08)$$
$${\rm [Fe/H]}=(1.44\pm0.03) \ (V-K) - (4.81\pm0.07)$$
Both relationships are in agreement with the empirical ones within the
1$\sigma$ errors. Analogue relationships obtained from the $\eta$=0.4 isochrones 
differ by much more than the 1$\sigma$ errors.

\end{document}